\documentclass[letterpaper,twocolumn]{emulateapj}
\usepackage[german,english,american]{babel} 
\usepackage{enumitem}
\usepackage{graphicx}
\usepackage{natbib}
\usepackage{color}
\usepackage{ifthen}

\definecolor{grey}{rgb}{0.7,0.7,0.7}

\def\lesssim{\mathrel{\hbox{\rlap{\hbox{\lower4pt\hbox{$\sim$}}}\hbox{$<$}}}}
\def\gtrsim{\mathrel{\hbox{\rlap{\hbox{\lower4pt\hbox{$\sim$}}}\hbox{$>$}}}}
\newcommand{\cir}{Circinus X-1\ }

\newlength{\figwidth}
\newcommand{\pasa}{Pub.Astr.Soc.Austr.}

\begin{document}

\title{Lord of the Rings: A Kinematic Distance to Circinus X-1 from a
  Giant X-Ray Light Echo}

\author{S. Heinz\altaffilmark{1}}
\email{heinzs@astro.wisc.edu}
\author{M. Burton\altaffilmark{2}}
\author{C. Braiding\altaffilmark{2}}
\author{W.N. Brandt\altaffilmark{3,4,5}}
\author{P.G. Jonker\altaffilmark{6,7,8}}
\author{P. Sell\altaffilmark{9}}
\author{R.P. Fender\altaffilmark{10}}
\author{M.A. Nowak\altaffilmark{11}}
\author{N.S. Schulz\altaffilmark{11}}

\altaffiltext{1}{Department of Astronomy, University of Wisconsin-Madison,
  Madison, WI 53706, USA} 
\altaffiltext{2}{School of Physics, University of New South Wales,
  Sydney, NSW 2052, Australia}
\altaffiltext{3}{Department of Astronomy \& Astrophysics, The
  Pennsylvania State University, University Park, PA 16802, USA}
\altaffiltext{4}{Institute for Gravitation and the Cosmos, The Pennsylvania
 State University, University Park, PA 16802, USA}
\altaffiltext{5}{Department of Physics, The Pennsylvania State University,
 University Park, PA 16802, USA}
\altaffiltext{6}{SRON, Netherlands Institute for Space Research, 3584
  CA, Utrecht, the Netherlands}
\altaffiltext{7}{Department of Astrophysics/IMAPP, Radboud University
  Nijmegen, 6500 GL, Nijmegen, The Netherlands}
\altaffiltext{8}{Harvard--Smithsonian Center for Astrophysics,
  Cambridge, MA~02138, USA}
\altaffiltext{9}{Physics Department,Texas Technical University,
  Lubbock, TX 79409, USA}
\altaffiltext{10}{Department of Astronomy, University of Oxford,
  Astrophysics, Oxford OX1 3RH, UK} 
\altaffiltext{11}{Kavli Institute for Astrophysics and Space Research,
  Massachusetts Institute of Technology, Cambridge, MA 02139, USA}

\begin{abstract}
  Circinus X-1 exhibited a bright X-ray flare in late 2013. Follow-up
  observations with {\em Chandra} and {\em XMM-Newton} from 40 to 80
  days after the flare reveal a bright X-ray light echo in the form of
  four well-defined rings with radii from 5 to 13 arcminutes, growing
  in radius with time.  The large fluence of the flare and the large
  column density of interstellar dust towards Circinus X-1 make this
  the largest and brightest set of rings from an X-ray light echo
  observed to date.  By deconvolving the radial intensity profile of
  the echo with the MAXI X-ray lightcurve of the flare we reconstruct
  the dust distribution towards Circinus X-1 into four distinct dust
  concentrations. By comparing the peak in scattering intensity with
  the peak intensity in CO maps of molecular clouds from the Mopra
  Southern Galactic Plane CO Survey we identify the two innermost
  rings with clouds at radial velocity $\sim-74\,{\rm km\,s^{-1}}$ and
  $\sim-81\,{\rm km\,s^{-1}}$, respectively.  We identify a prominent
  band of foreground photoelectric absorption with a lane of CO gas at
  $\sim-32\,{\rm km\,s^{-1}}$.  From the association of the rings with
  individual CO clouds we determine the kinematic distance to \cir to
  be $D_{\rm Cir X-1} = 9.4^{+0.8}_{-1.0}\,{\rm kpc}$.  This distance
  rules out earlier claims of a distance around $4\,{\rm kpc}$,
  implies that \cir is a frequent super-Eddington source, and places a
  lower limit of $\Gamma \gtrsim 22$ on the Lorentz factor and an
  upper limit of $\theta_{\rm jet} \lesssim 3^{\circ}$ on the jet
  viewing angle.
\end{abstract}


\section{Introduction}

\subsection{X-ray Dust Scattering Light Echoes}
X-rays from bright point sources are affected by scattering off
interstellar dust grains as they travel towards the observer.  For
sufficiently large dust column densities, a significant fraction of
the flux can be scattered into an arcminute-sized, soft X-ray halo.
Dust scattering halos have long been used to study the properties of
the interstellar medium and, in some cases, to constrain the
properties of the X-ray source
\citep[e.g.][]{mathis:91,predehl:95,corrales:13}.

When the X-ray source is variable, the observable time-delay
experienced by the scattered X-ray photons relative to the un-scattered
X-rays can provide an even more powerful probe of the ISM than a
constant X-ray halo.  In this case, the dust scattering halo will show
temporal variations in intensity that can be cross-correlated with the
light curve to constrain, for example, the distance to the dust clouds
along the line-of-sight \citep[e.g.,][]{xiang:11}.

A special situation arises when the X-ray source exhibits a strong and
temporally well-defined flare (as observed, for example, in gamma-ray
bursts and magnetars). In this case, the scattered X-ray signal is not
a temporal and spatial average of the source flux history and the
Galactic dust distribution, but a well-defined light echo in the form
of distinct rings of X-rays that increase in radius as time passes,
given the longer and longer time delays \citep{vianello:07}.

\begin{figure}[t]
  \center\resizebox{\columnwidth}{!}{\includegraphics{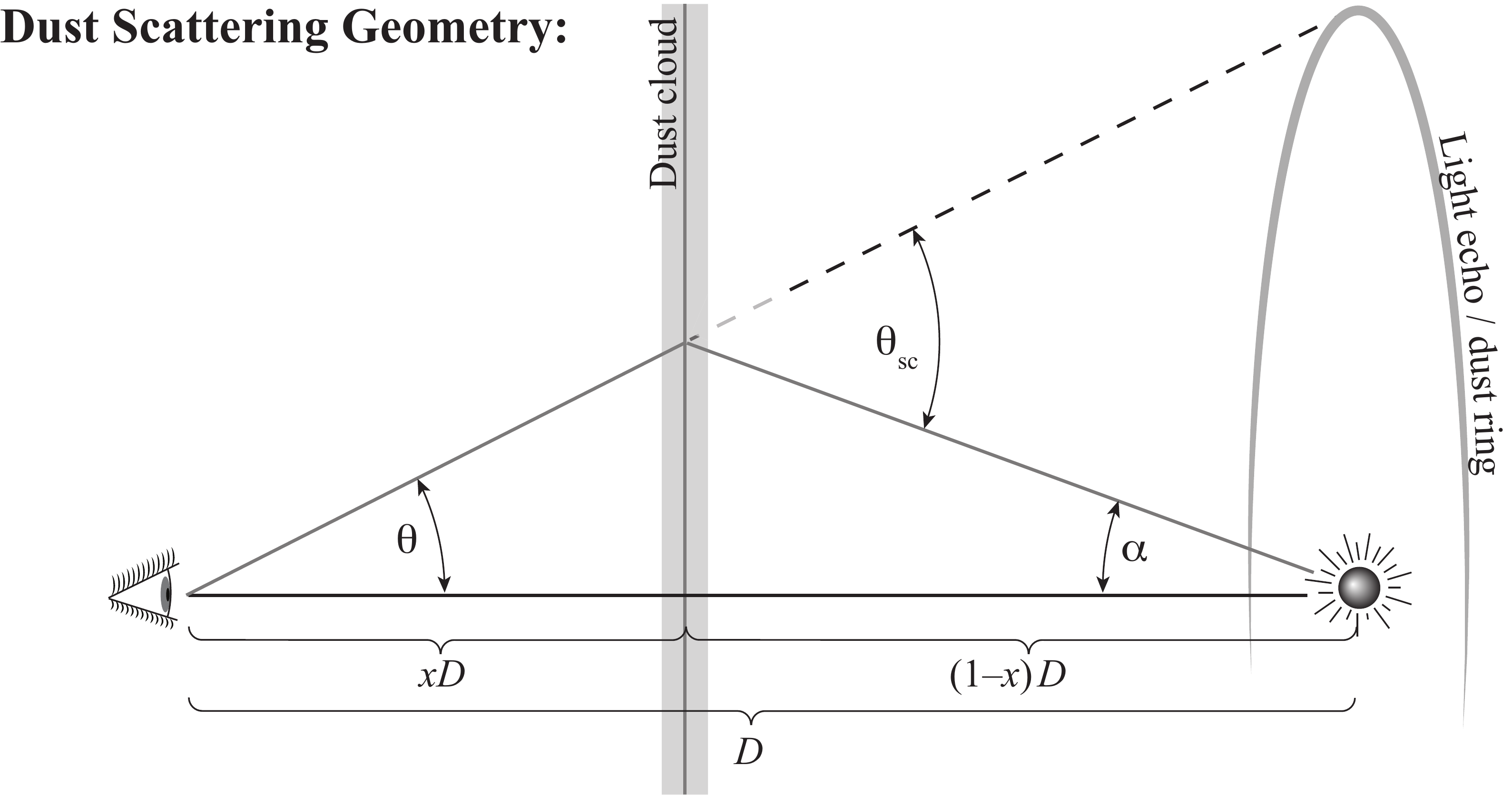}}
  \caption{Cartoon of dust scattering geometry and nomenclature used
    throughout this paper.  X-rays from a source at distance $D$
    scatter off a single dust layer at distance $xD$ (such that the
    fractional distance is $x$). The observed angle of the scattered
    X-rays is $\theta$, while the true scattering angle is
    $\theta_{\rm sc} =\theta + \alpha$. The scattered X-rays travel an
    additional path length of $\Delta D= \Delta D_{1} + \Delta
    D_{2}$.}\label{fig:geometry}\vspace*{6pt}
\end{figure}

The scattering geometry of X-rays reflected by an interstellar dust
cloud is illustrated in Fig.~\ref{fig:geometry}. For a source distance
$D$ and a layer of interstellar dust at distance $xD$ from the
observer (such that the fractional distance to the dust layer is $x$),
the scattered X-rays must travel an additional distance $\Delta D$
which depends on the observed off-axis angle $\theta$
\citep[e.g.][]{mathis:91,xiang:11}:
\begin{eqnarray}
  \Delta D & = & 
  xD\left(\frac{1}{\cos({\theta})} - 1\right) + \left(1 -
    x\right)D\left(\frac{1}{\cos({\alpha})} - 1\right) \nonumber\\
  & \approx & \frac{xD\theta^2 + (1-x)D\alpha^2}{2}
  = \frac{xD\theta^2}{2\left(1-x\right)}
\end{eqnarray}
where we have used the small-angle approximation. The scattered signal
will arrive with a time delay of
\begin{equation}
  \label{eq:deltat}
  \Delta t = \frac{\Delta D}{c} = \frac{xD\theta^2}{2c(1-x)}
\end{equation}

For a well-defined (short) flare of X-rays, the scattered flare signal
propagates outward from the source in an annulus of angle
\begin{equation}
 \label{eq:theta}
  \theta=\sqrt{\frac{2 c \Delta t (1 - x)}{xD}}
\end{equation}
For a point source X-ray flux $F(t)$ scattered by a thin sheet of dust
of column density $N_{\rm H}$, the X-ray intensity $I_{\nu}$ of the
dust scattering annulus at angle $\theta$ observed at time $t_{\rm
  obs}$ is given by \citep{mathis:91}
\begin{equation}
  \label{eq:intensity}
  I_{\nu}(\theta,t_{\rm obs})=N_{\rm H}\frac{F_{\nu}(t_{\rm
      obs}-\Delta t_{\theta})}{\left(1 - x\right)^2}\frac{d\sigma_{\nu}}{d\Omega}
\end{equation}
where $\Delta t_{\theta}$ is given by eq.~(\ref{eq:deltat}) and
$d\sigma_{\nu}/d\Omega$ is the differential dust scattering cross
section per Hydrogen atom, integrated over the grain size distribution
\citep[e.g.][]{mathis:91,draine:03}.  In the following, we will refer
to the quantity $N_{\rm H}d\sigma_{\nu}/d\Omega$ as the dust
scattering depth.

For a flare of finite duration scattering off a dust layer of finite
thickness, the scattered signal will consist of a set of rings that
reflect the different delay times for the different parts of the
lightcurve relative to the time of the observation and the
distribution of dust along the line of sight.\footnote{Because we will
  discuss delay times of several months, the much shorter duration of
  the X-ray observations may be neglected in the following
  discussion.}

\begin{figure}[t]
  \center\resizebox{\columnwidth}{!}{\includegraphics{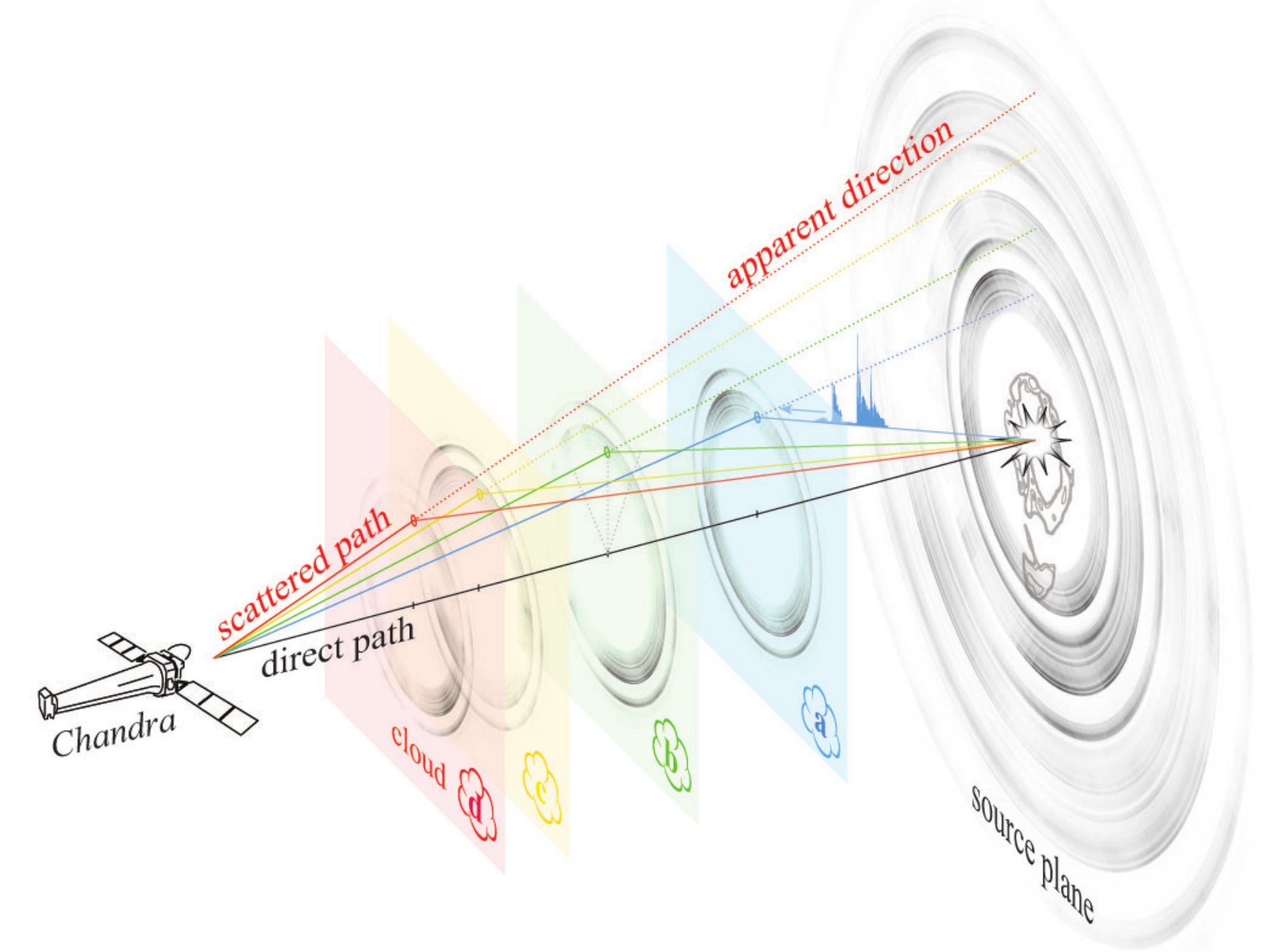}}
  \caption{Cartoon of the X-ray dust-scattering echo analyzed in this
    paper: A bright flare from a point source (center of the source
    plane) propagates towards the observer. Dust in dense interstellar
    clouds between observer and source scatters X-rays towards the
    observer, producing a light echo in the form of well defined
    rings. In this example, four clouds [a]-[d] shown in blue, green,
    yellow, and red, are located at different distances, illuminated
    by a bright flare (blue lightcurve) and producing four well
    defined rings in the image/source
    plane. \vspace*{6pt}}\label{fig:cartoon}
\end{figure}

In the presence of multiple well-defined scattering layers at
different fractional distances $x$ (with thickness $l_{\rm dust} \ll
D$), each layer will generate a separate set of annuli. At a given
time delay, the annuli of a fixed, short segment of the lightcurve are
defined by the intersections of an ellipsoid of constant path length
and each scattering layer. The superposition of the annuli from all
dust layers (clouds) and the different parts of the flare light curve
will create a set of (partially overlapping) rings. This situation is
illustrated in Fig.~\ref{fig:cartoon}.

If the lightcurve of the X-ray flare that created the echo is known,
an X-ray image of such a set of light-echo rings with sufficiently
high angular resolution and signal-to-noise may be de-composed into
contributions from a set of distinct dust layers at different
distances. Such observations require both high angular resolution and
sufficiently large collecting area and low background to detect the
rings. The {\em Chandra} X-ray observatory is ideally suited for such
observations.

\subsection{Dust Echoes of Galactic X-ray Sources}
X-ray rings generated by dust scattering light echoes are extremely
rare.  To date, rings have been reported from five gamma-ray bursts
\citep[][and references therein]{vianello:07}.  For Galactic X-ray
sources, observations of light echoes have been even more elusive. The
soft gamma-ray repeater 1E 1547.0-5408 was the first Galactic X-ray
source to show a clear set of bright X-ray light echo rings
\citep{tiengo:10}. Two other sources have also been reported to show
light echo rings: the soft gamma-ray repeater SGR 1806-20
\citep{svirski:11} and the fast X-ray transient IGR J17544-2619
\citep{mao:14}.  However, the fluence of the initial X-ray flare in
the latter two cases was too low to produce clearly resolved,
arcminute-scale rings and both detections are tentative.  In addition,
\citet{mccollough:13} found repeated scattering echoes from a single
ISM cloud towards Cygnus X-3.

\begin{figure}[t]
  \center\resizebox{\columnwidth}{!}{\includegraphics{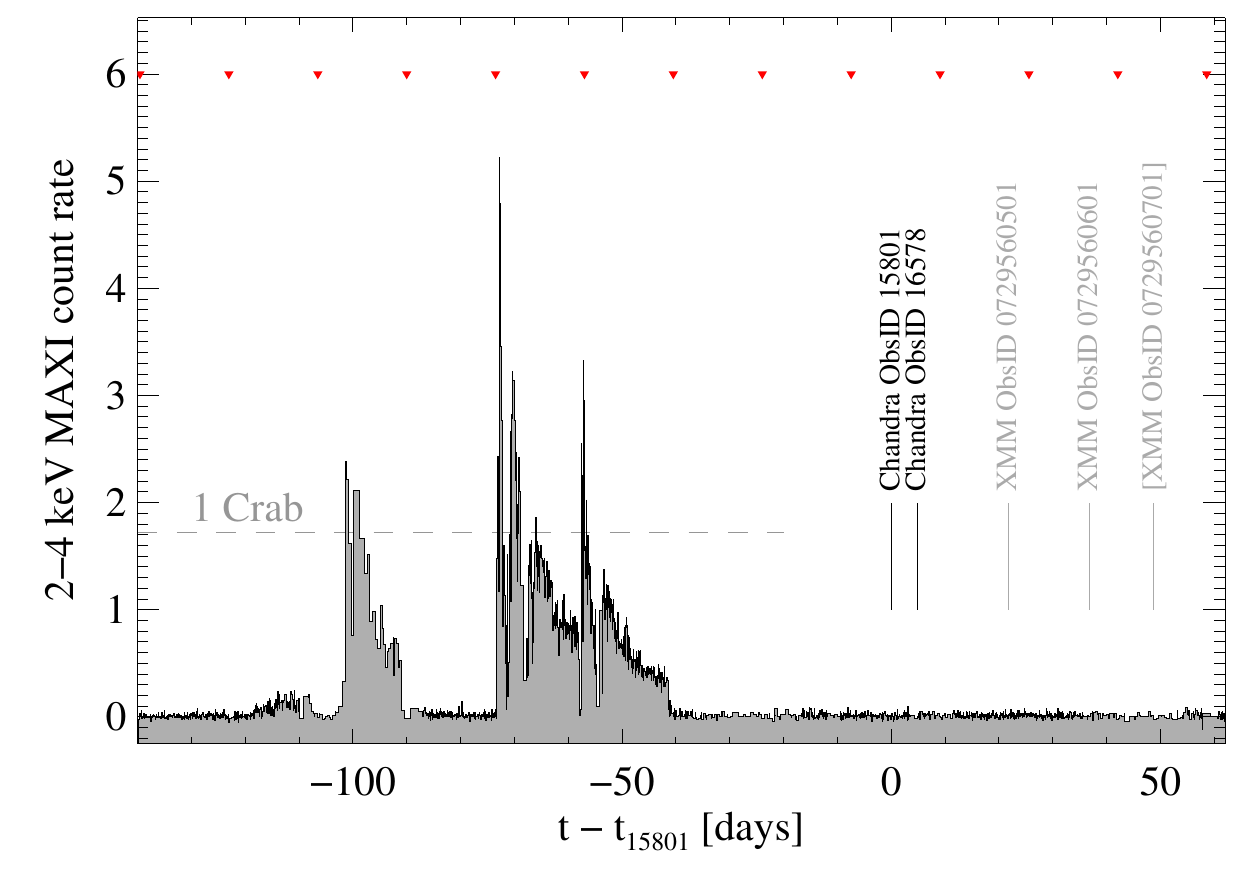}}
  \caption{MAXI 2-4 keV lightcurve of Circinus X-1 at the time of the
    2013 flare. Also shown are the median-times of our two {\em
      Chandra} and three {\em XMM} observations ({\em XMM} ObsID
    0729560701 was not used in this paper).  $T_{0}$ was chosen to
    correspond to the time of our first {\rm Chandra} observation at
    ($T_{0}=$MJD$_{\rm 15801}$=56683). Red diamond symbols
    indicate periastron at orbital phase zero
    \citep{nicolson:07}. \vspace*{6pt}}\label{fig:maxi_lightcurve}
\end{figure}

In the case of GRBs, the source is known to be at cosmological
distances, which makes the scattering geometry particularly simple,
since the source is effectively at infinite distance and the
scattering angle is equal to the observed ring angle on the sky, which
simplifies eq.~(\ref{eq:theta}) to
\begin{equation}
  \theta_{\rm GRB echo} = \sqrt{\frac{2c\Delta t}{D_{\rm dust}}}
    \label{eq:theta_grb}
\end{equation}
An observed ring radius $\theta_{\rm GRB echo}$ and delay time
$\Delta t$ therefore unambiguously determines the distance to the
dust.

However, in the case of Galactic sources, eq.~(\ref{eq:theta}) for
$\theta(\Delta t, D, x)$ cannot be inverted to solve for the distance
to either the dust or the source individually, unless the distance to
either the source or the dust is known apriori. {\em If} a particular
ring can be identified with a dust cloud of {\em known} distance $xD$,
the distance to the source can be determined from the known delay time
$\Delta t$.

\subsection{Circinus X-1}
\label{sec:cir}
\cir is a highly variable X-ray binary that has often been
characterized as erratic. It has been difficult to classify in terms
of canonical X-ray binary schemes, combining properties of young and
old X-ray binaries
\citep{stewart:91,oosterbroek:95,jonker:07,calvelo:12b}. The presence
of type I X-ray bursts \citep{tennant:86,linares:10} identifies the
compact object in the source as a neutron star with a low magnetic
field, supported by the presence of jets and the lack of X-ray or
radio pulses
\citep{stewart:93,fender:04c,tudose:06,heinz:07,soleri:09,sell:10}.

\begin{figure}[tbp ]
  \center\resizebox{\columnwidth}{!}{\includegraphics{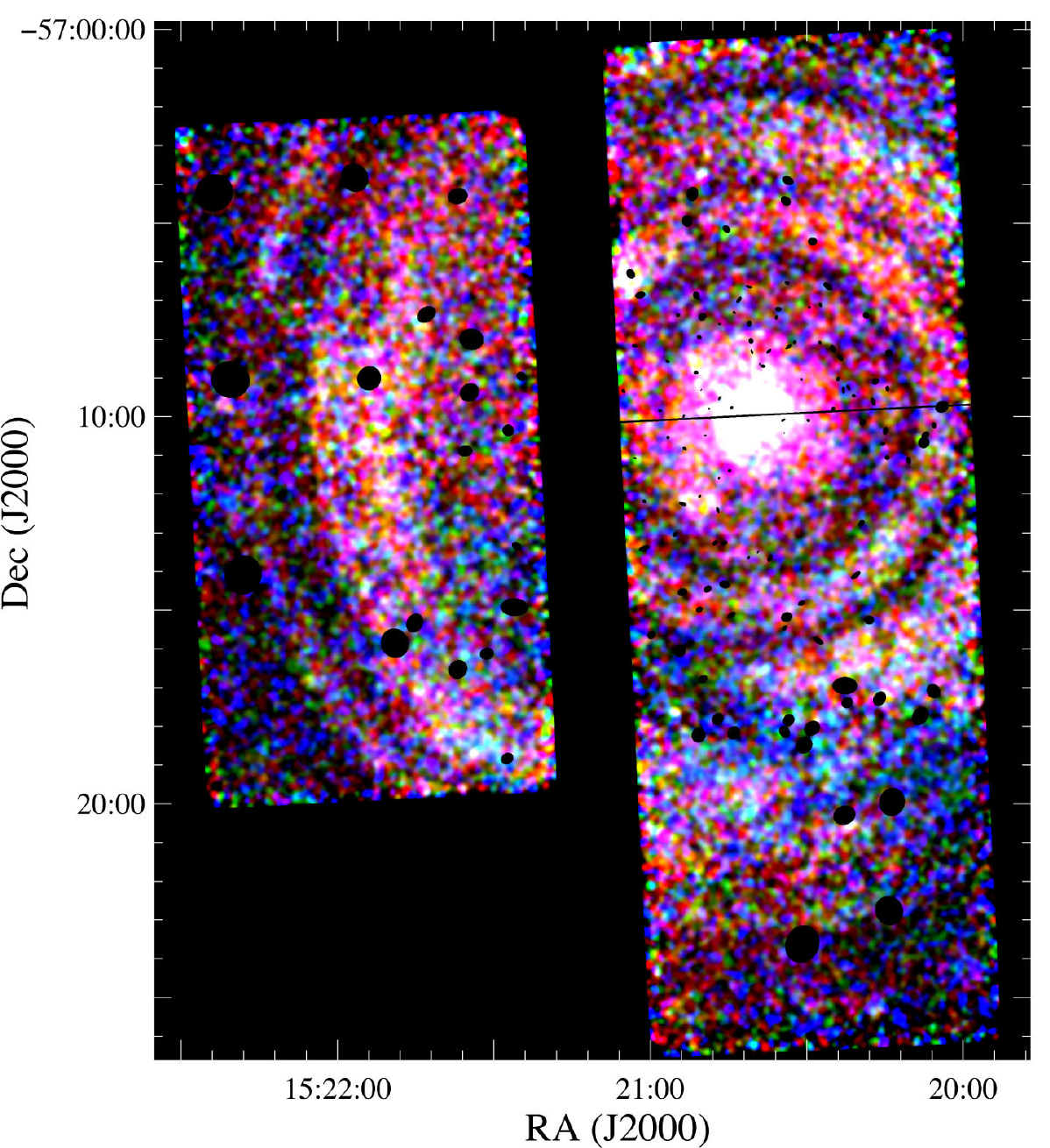}}
  \caption{Exposure-corrected image of {\em Chandra} ObsID 15801,
    smoothed with a FWHM=10.5" Gaussian. Red: 1-2 keV, green: 2-3 keV,
    blue: 3-5 keV. Point sources and the read streak produced by \cir
    were removed. Four separate rings of dust scattering echoes are
    visible in this image. \vspace*{6pt}}
\label{fig:color_chandra_image}
\end{figure}
 
The transient nature and the low magnetic field suggested a
classification of \cir as an old, low-mass X-ray binary with an
evolved companion star \citep[e.g.][]{shirey:96}. This interpretation
was in conflict with the observed orbital evolution time of order
$P/\dot{P}\,\sim\,3,000$ years \citep{parkinson:03,clarkson:04} to
$P/\dot{P}\,\sim\,20,000$ years \citep{nicolson:07} and the possible
identification of an A-B type super-giant companion star
\citep{jonker:07}, suggesting a much younger system age and a likely
massive companion star.  The orbital period of the system is 16.5
days, with a likely eccentricity around 0.45 \citep{jonker:07}

The conflicting identifications were resolved with the discovery of an
arcminute scale supernova remnant in both X-ray and radio
observations \citep{heinz:13}, placing an upper limit of $\tau <
4,600\,{\rm years} \,(D_{\rm Cir}/8\,{\rm kpc})$ on the age of the
system, where $D_{\rm Cir}$ is the distance to the source.  This upper
limit makes \cir the youngest known X-ray binary and an important test
case for the study of both neutron star formation and orbital
evolution in X-ray binaries.

The distance to \cir is highly uncertain.  As a southern source deeply
embedded in the Galactic disk, most standard methods of distance
determination are not available.  It is also out of the range of VLBI
parallax measurements.  Distance estimates to \cir range from $4\,{\rm
  kpc}$ \citep{iaria:05} to $11\,{\rm kpc}$ \citep{stewart:91}. A
likely range of $D=8-10.5\,{\rm kpc}$ was proposed by
\citet{jonker:04} based on the radius-expansion burst method, using
properties of the observed type I X-ray bursts.  Because important
binary properties depend on the distance to the source (such as the
Eddington fraction of the source in outburst), a more accurate
determination of the distance is critical for a more complete
understanding of this important source.

\cir is a highly variable source, spanning over four orders of
magnitude in luminosity from its quiescent flux to its brightest
flares.  While the source was consistently bright during the 1990's,
it underwent a secular decline in flux starting in about 2000 and has
spent the past 10 years below the detection thresholds of the {\em
  Rossi X-ray Timing Explorer} All Sky Monitor and the MAXI all sky
monitor aboard the {\em International Space Station}
\citep{matsuoka:09}\footnote{The ASM was one of three instruments
  aboard the {\em Rossi X-ray Timing Explorer}; MAXI is an experiment
  aboard the {\em International Space Station}; both have only one
  mode of operation.}, with the exception of sporadic near-Eddington
flares, sometimes in excess of one Crab in flux (throughout the paper
we will refer to the Eddington limit of $L_{\rm Edd,NS} \sim 1.8\times
10^{38}\,{\rm ergs\,s^{-1}}$ for a $1.4\,{\rm M}_{\odot}$ neutron
star). These flares typically occur within individual binary orbits
and are characterized by very rapid rises in flux near periastron of
the binary orbit and often rapid flux decays at the end of the
orbit. For example, during the five years it has been monitored by
MAXI, \cir exhibited ten binary orbits with peak 2-10 keV fluxes in
excess of $\sim 10^{-8}\,{\rm ergs\,cm^{-2}\,s^{-1}}$ and five orbits
with peak flux at or above one Crab.

The large dynamic range of the source flux, the well-defined duration
of the X-ray flares, and the low mean flux during the quiescent
periods over the past decades make \cir an ideal source for
observations of X-ray light echoes.  The source is located in the
Galactic plane, at Galactic coordinates
$l=322.12^{\circ}, b=0.04^{\circ}$. The large neutral hydrogen column
density of $N_{\rm H}\sim 2\times 10^{22}\,{\rm cm^{-2}}$, inferred
from photo-electric absorption of the X-ray spectra of both the
neutron star \citep[e.g.][]{brandt:00} and the supernova remnant
\citep{heinz:13}, further increases the likelihood to observe light
echoes.

In late 2013, \cir underwent an extremely bright flare, preceded and
followed by long periods of very low X-ray flux.  In this paper, we
report the detection of the brightest and largest set of X-ray dust
scattering rings observed to date, generated by the light echo from
the 2013 flare. The spatial variation of the rings allows us to
identify the interstellar clouds responsible for the different rings,
which determines the distance to the X-ray source, Circinus X-1, to
roughly 10\% accuracy.

\begin{figure}[t]
  \center\resizebox{\columnwidth}{!}{\includegraphics{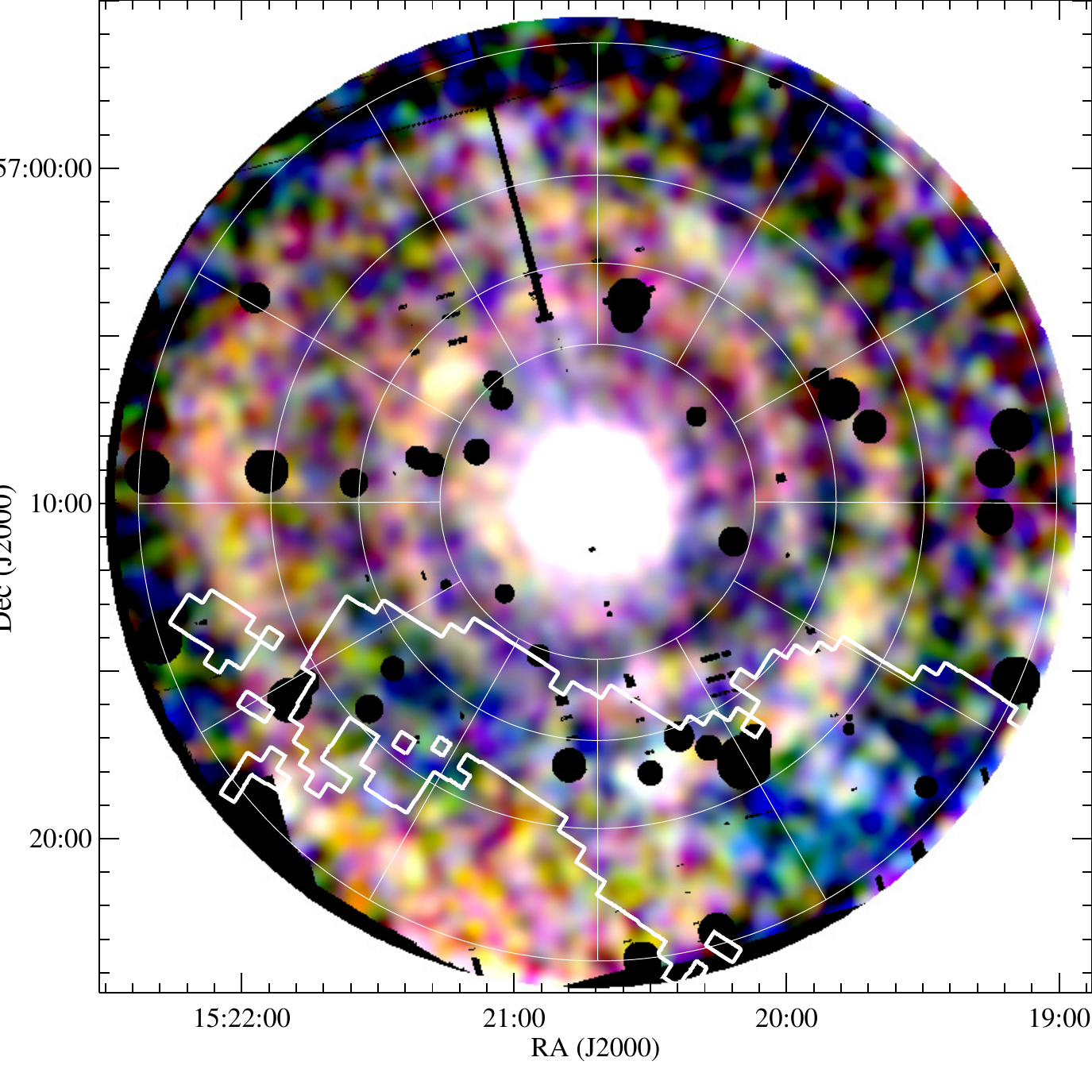}}
  \caption{Exposure-corrected and background-subtracted image of {\em
      XMM-Newton} ObsID 0729560501 (combining {\em EPIC} MOS and PN
    data), smoothed with a FWHM=35" Gaussian. Colors represent the
    same energy bands as in Fig.~\ref{fig:color_chandra_image}. Point
    sources have been removed.  The over-plotted white-dashed line
    corresponds to the $15 {\rm K\,km\,s^{-1}}$ contour of the $-33.6$
    to $-29.6\,{\rm km\,s^{-1}}$ $^{12}$CO image shown in
    Fig.~\ref{fig:mopra_6}. \vspace*{6pt} \label{fig:color_xmm_image}}
\end{figure}

 We will describe the observations and data analysis in
\S\ref{sec:observations}. Section \ref{sec:analysis} describes the
analysis of the X-ray, CO, and HI data and the procedure for
deconvolving the X-ray signal into the line-of-sight dust
distribution. In \S\ref{sec:discussion}, we derive a new distance to
Circinus X-1 and discuss the consequences for the physics of this
unique X-ray binary and for the physics of dust scattering. Section
\ref{sec:conclusions} briefly summarizes our findings.

\section{Observations and Data Reduction}
\label{sec:observations}
\subsection{X-Ray Observations}

Circinus X-1 was observed with {\em Chandra}, {\em XMM-Newton}, and
{\em Swift} in early 2014 after a bright X-ray flare of the source in
late 2013.  Figure \ref{fig:maxi_lightcurve} shows the time of the
{\em Chandra} and {\em XMM-Newton} observations overlaid on the MAXI
lightcurve of the source. The source exhibited a bright flare during
three binary orbits approximately 50 to 100 days before the first {\em
  Chandra} observation \citep{asai:14}, with a quiescent binary orbit
interspersed during the first half of the flare.  The source flare
reached fluxes in excess of one Crab, with a total 2-4 keV fluence of
${\mathcal F}_{2-4}=0.025\,{\rm ergs\,cm^{-2}}$.  The lightcurve of
the flare has three characteristic peaks and a pronounced gap.  This
is the most energetic flare \cir exhibited since it entered its
long-term quiescent state in 2006, and is comparable in peak flux to
the brightest flares the source has undergone even during its
long-term outburst.

Before and during the {\em Chandra} campaign, {\em Swift} observed the
source twelve times for a total exposure of 12 ksec in order to safely
schedule the {\em Chandra} observations. The history of bright flares
by Circinus X-1, which can exceed the ACIS dose limit even during
moderate exposures presents a significant scheduling challenge, and
frequent monitoring of the source with Swift was used to confirm that
the source was safely below the maximum flux allowable under {\em
  Chandra} instrument safety rules.  The short total exposure time and
the fact that the exposures were spread over a time window of 11 days
make the {\em Swift} data unsuitable for the analysis presented in
this paper.

\begin{figure}[t]
  \center\resizebox{\columnwidth}{!}{\includegraphics{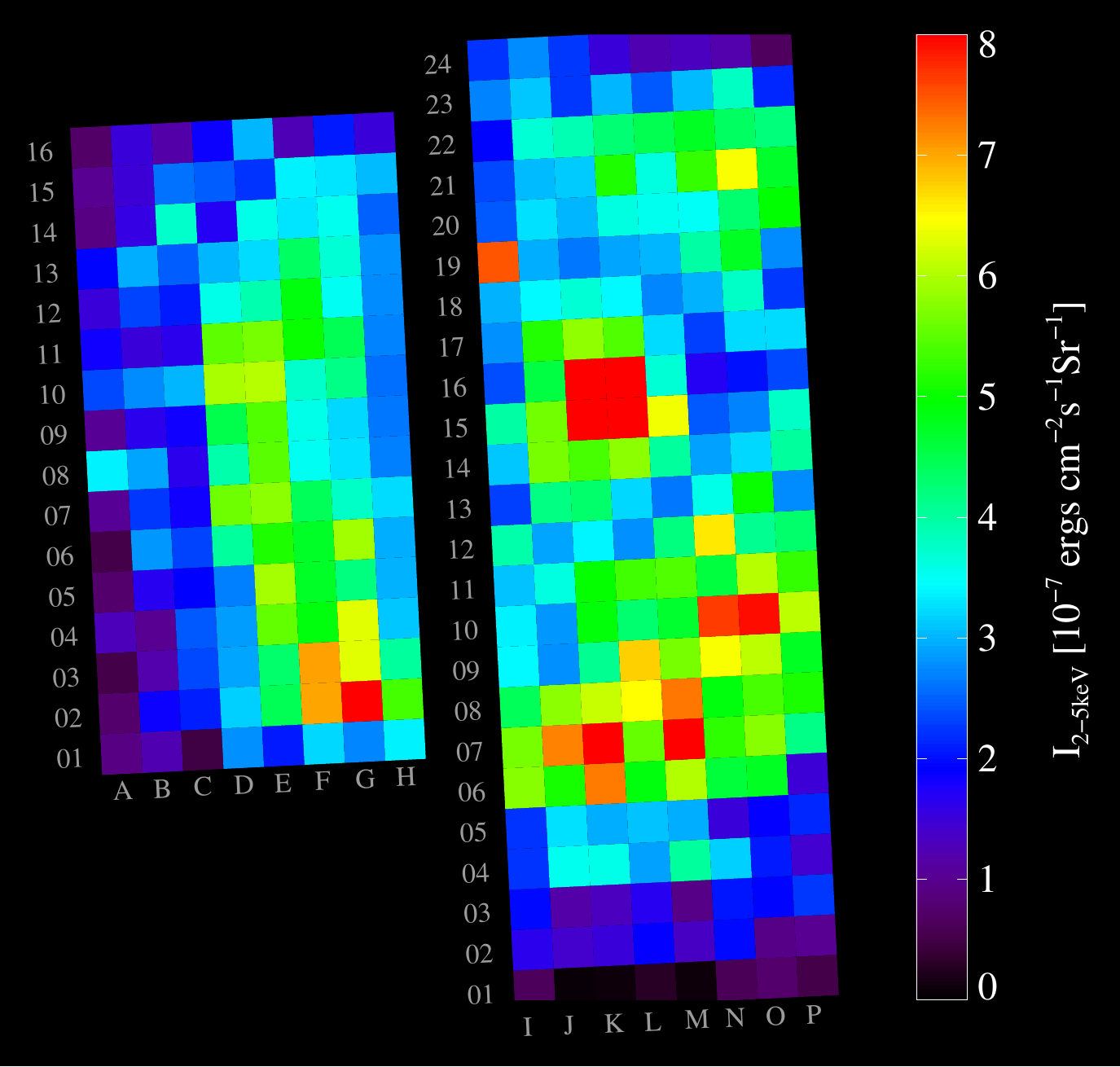}}
  \caption{Un-absorbed 2-5 keV scattering intensity in units of
    $10^{-7}\,{\rm ergs\,cm^{-2}\,s^{-1}\,Sr^{-1}}$ across the {\em
      Chandra} field-of-view, derived from spectral fits to {\em
      Chandra} ObsID 15801 on a rectangular grid of 8$\times$8 squares
    per CCD, labeled by row and column from 1 through 24 and A through
    P, respectively.  Photo-electric absorption was removed to isolate
    the dust-scattered intensity. \vspace*{6pt}}
  \label{fig:chandra_norms}
\end{figure}

\subsubsection{Chandra Observations}
In the 3.5 binary orbits preceding the first {\em Chandra}
observation, as well as during the months prior to the flare, the
source was quiescent, below the MAXI detection threshold. We observed
the source on two occasions with {\em Chandra}, on January 25, 2014
for 125 ksec, and on January 31, 2014 for 55 ksec, listed as ObsID
15801 and 16578, respectively. The point source was at very low flux
levels during both observations. Circinus X-1 was placed on the ACIS
S3 chip.

Data were pipeline processed using {\tt CIAO} software version
4.6.2. Point sources were identified using the {\tt wavdetect}
\citep{freeman:02} task and ObsID 16578 was reprojected to match the
astrometry of ObsID 15801. For comparison and analysis purposes, we
also reprocessed ObsID 10062 (2009) with {\tt CIAO} 4.6.2 and
reprojected it to match the astrometry of ObsID 15801.

We prepared blank background images following the standard {\tt CIAO}
thread \citep[see also][]{hickox:06}, matching the hard ($>$10 keV)
X-ray spectrum of the background file for each chip.

 Figure \ref{fig:color_chandra_image} shows an exposure-corrected,
background-subtracted three-color image of the full ACIS field-of view
captured by the observation (ACIS chips 2,3,6,7, and 8 were active
during the observation), where red, green, and blue correspond to the
1-2, 2-3, and 3-5 keV bands, respectively. Point sources were
identified using {\tt wavdetect} in ObsIDs 10062, 15801, and 16578,
source lists were merged and point sources were removed from imaging
and spectral analysis. The read streak was removed and the image was
smoothed with a 10.5" full-width-half-max (FWHM) Gaussian in all three
bands.

The image shows the X-ray binary point source, the X-ray jets (both
over-exposed in the center of the image), and the supernova remnant in
the central part of the image around the source position at
15:20:40.9,-57:10:00.1 (J2000). 

The image also clearly shows at least three bright rings that are
concentric on the point source. The first ring spans from 4.2 to 5.7
arcminutes in radius, the second ring from 6.1 to 8.2 arcminutes, and the
third from 8.3 to 11.4 arcminutes in radius, predominantly covered by the
Eastern chips 2 and 3. We will refer to these rings as rings [a], [b],
and [c] from the inside out, respectively.  An additional ring-like
excess is visible at approximately 13 arcminutes in radius, which we
will refer to as ring [d].

As we will lay out in detail in \S\ref{sec:discussion}, we interpret
these rings as the dust-scattering echo from the bright flare in
Oct.-Dec. of 2013, with each ring corresponding to a distinct
concentration of dust along the line-of-sight to Circinus X-1. A
continuous dust distribution would not produce the distinct, sharp set
of rings observed by {\em Chandra}.

\begin{figure}[t]
  \center\resizebox{\columnwidth}{!}{\includegraphics{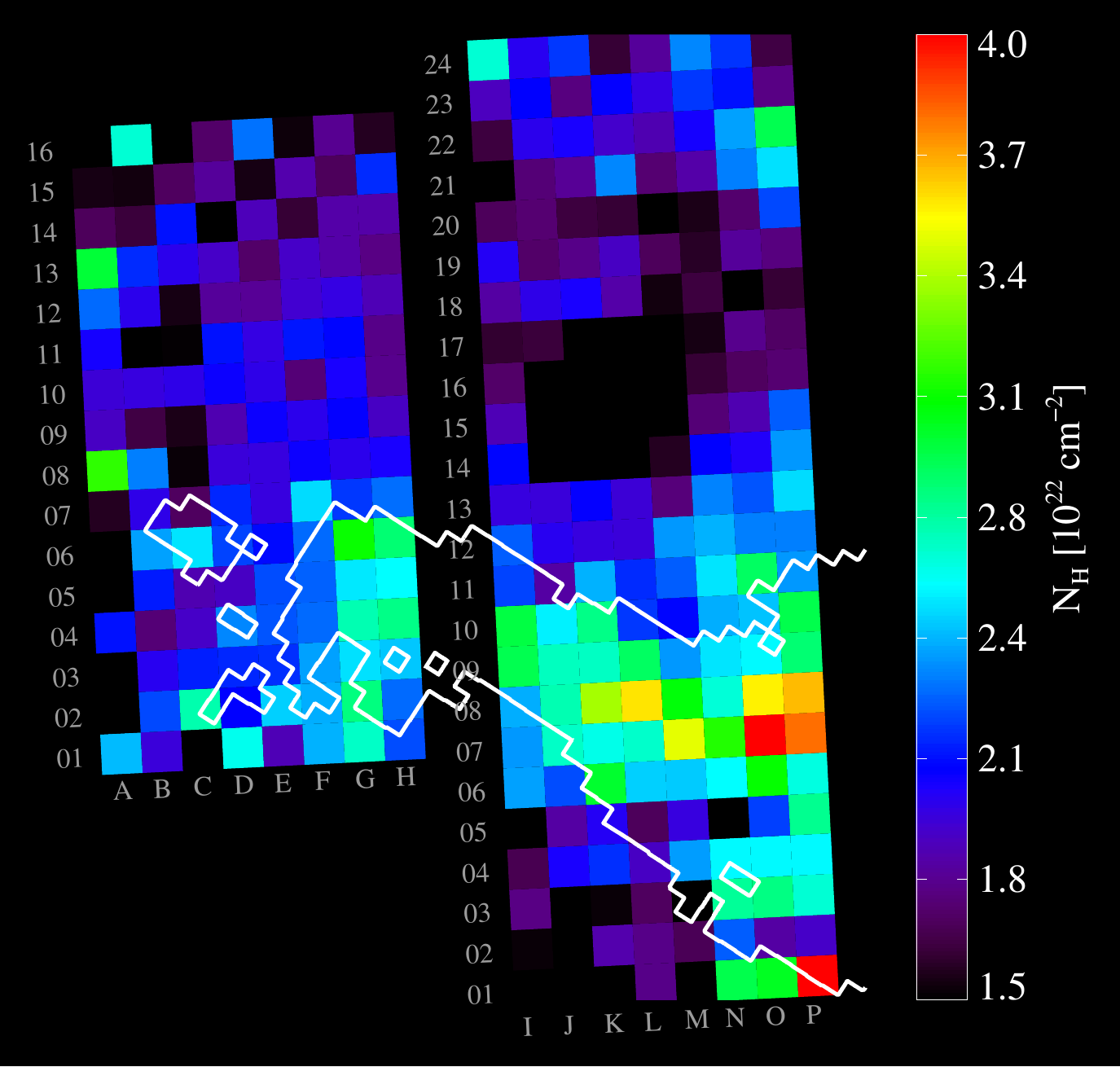}}
  \caption{Distribution of photo-electric absorption column $N_{\rm
      H}$ derived from the spectral fits shown in
    Fig.~\ref{fig:chandra_norms}. A clear excess of absorption is
    found in the lower half of the image. Over-plotted is the
    $15\,{\rm K\,km\,s^{-1}}$ contour of the $-33.6$ to $-29.6\,{\rm
      km\,s^{-1}}$ $^{12}$CO image shown in
    Fig.~\ref{fig:mopra_6}. \vspace*{6pt}}
  \label{fig:chandra_columns}
\end{figure}

The rings are also clearly visible in ObsID 16578, despite the shorter
exposure and the resulting lower signal-to-noise.  In ObsID 16578, the
rings appear at $\sim$4\% larger radii, consistent with the
expectation of a dust echo moving outward in radius [see also
\S\ref{sec:profiles} and eq.~(\ref{eq:ring_increase})].  The rings are
easily discernible by eye in the energy bands from 1 to 5 keV.

Even though the outer rings are not fully covered by the {\em Chandra}
field of view (abbreviated as FOV in the following), it is clear that
the rings are not uniform in brightness as a function of azimuthal
angle. There are clear intensity peaks at [15:21:00,-57:06:30] in the
inner ring [a] and at [15:20:20,-57:16:00] in ring [b] of ObsID
15801. Generally, ring [c] appears brighter on theSouth- Eastern side
of the image.

The deviation from axi-symmetry observed here differs from almost all
other observations of dust scattering signatures, which typically
appear to be very uniform in azimuth.  For example, a detailed
investigation of the dust scattering halo of Cygnus X-2 found that the
profile deviated by only about 2\% from axi-symmetry
\citep{seward:13}.  The only comparable observational signature of
non-symmetric scattering features is the light echo off a bok globule
observed for Cygnus X-3 \citep{mccollough:13}

In addition to the light echo, a blue "lane" is visible across the
lower part of the image. As we will show below, this feature is the
result of photoelectric foreground absorption by an additional layer
of dust and gas closer to the observer than the clouds producing the
light echo.

\subsubsection{XMM-Newton Observations}

{\em XMM-Newton} observed the source on 17 February, 2014 (ObsID
0729560501), 4 March, 2014 (ObsID 0729560601), and 16 March, 2014
(ObsID 0729560701) for 30 ksec each. ObsID 0729560501 was not affected
by any background flares. The second and third observation were
strongly affected by background particle flares, eliminating the use
of PN data completely for ObsID 0729560601, and the effective exposure
for the MOS detectors after flare rejection was cut to approximately
15ksec in both cases. The resulting data are noisy, suggesting that
additional background contamination is present, making the data not
usable for detailed, quantitative image processing. We restrict the
discussion in this paper to data set 0729560501 (with the exception of
Fig.~\ref{fig:mopra_1}).

\begin{figure}[t]
  \center\resizebox{\columnwidth}{!}{\includegraphics{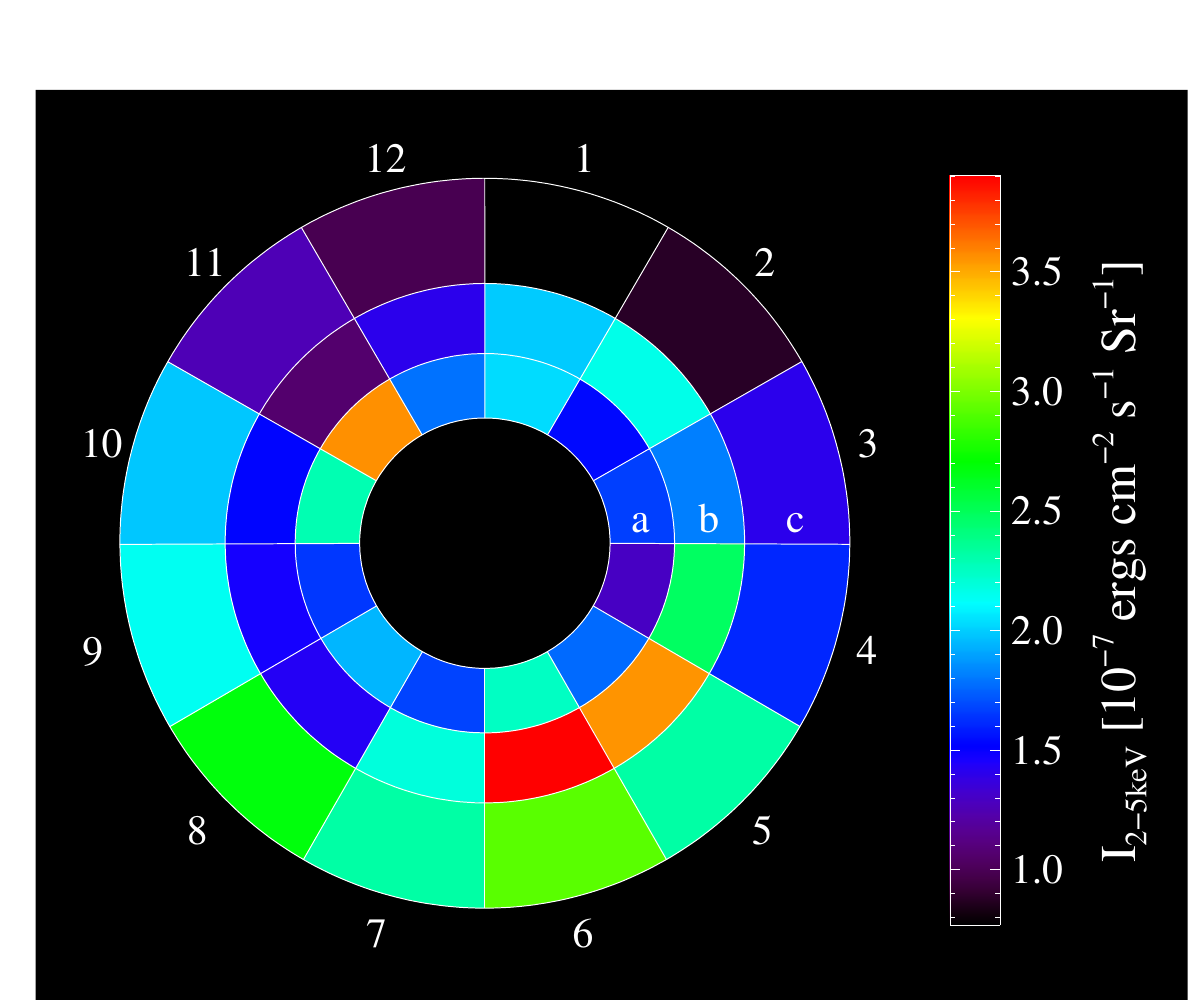}}
  \caption{Un-absorbed excess 2-5 keV scattering intensity in units of
    ${\rm ergs\,cm^{-2}\,s^{-1}\,Sr^{-1}}$ across the {\em XMM}
    field-of-view, derived from spectral fits in annular sections
    identical to those plotted in Fig.~\ref{fig:color_xmm_image}.
    Photo-electric absorption was removed to isolate the
    dust-scattered intensity. Clear peaks in the surface brightness
    are located ring sections section [a11] and [b6]. \vspace*{6pt}}
  \label{fig:norms}
\end{figure}

MOS and PN data were reduced using {\tt SAS} software version
13.5. Data were prepared using the {\tt ESAS} package, following the
standard diffuse source analysis guidelines from the Diffuse Analysis
Cookbook\footnote{ftp://xmm.esac.esa.int/pub/xmm-esas/xmm-esas.pdf}.
In addition to point sources identified by the {\tt ESAS cheese} task,
point sources identified in the three {\em Chandra} exposures (10062,
15801, 16578) were removed before image and spectral extraction.  The
energy range from 1.4-1.6 keV around the prominent instrumental
Aluminum line at 1.5 keV was removed from image processing, while
spectral fits determined that the contribution from the Si line at
1.75 keV was sufficiently small to ignore in image processing. Proton
and solar-wind charge exchange maps were created using spectral fits
to the outer regions of the image and subtracted during image
processing.

Figure \ref{fig:color_xmm_image} shows a three color
exposure-corrected image of ObsID 0729560501 in the same bands as
Fig.~\ref{fig:color_chandra_image}, smoothed with a 35'' FWHM
Gaussian.  Despite the lower spatial resolution of {\em XMM-Newton},
the image clearly shows the three rings identified in the {\em
  Chandra} images. The median radii of the rings have increased by
$\sim$16\% relative to ObsID 15801, corresponding to the longer time
delay between the flare and the {\em XMM} observation.

Overlaid on the figure is a grid of three annuli covering the rings,
broken into twelve angular sections of 30$^{\circ}$ each. We will use
this grid in the discussion below to identify different areas in the
image and in corresponding CO maps of the FOV.  The rings are labeled
[a],[b], and [c] from inside out, and the sections are labeled 1-12 in
a clock-wise fashion. See \S\ref{sec:xray_spectra} and
Fig.~\ref{fig:norms} for further details on the grid and the
nomenclature used in this paper.

\subsection{CO Data}

We used CO data from the Mopra Southern Galactic Plane CO Survey,
described in Burton et al. 2013, with the particular data set used
here, for the Circinus X-1 region (l=322°), coming from the
forthcoming public data release in Braiding et al., 2015 (in
preparation).  In this data set emission from the J=1-0 line of the
three principal isotopologues of the CO molecule ($^{12}$CO,
$^{13}$CO, C$^{18}$O, at 115.27, 110.20 and 109.78 GHz, respectively)
has been measured at 0.6 arcmin, $0.1\,{\rm km\,s^{-1}}$ resolution
with the 22m diameter Mopra millimetre wave telescope, located by
Siding Spring Observatory in New South Wales in Australia.

\begin{figure}[t]
  \center\resizebox{\columnwidth}{!}{\includegraphics{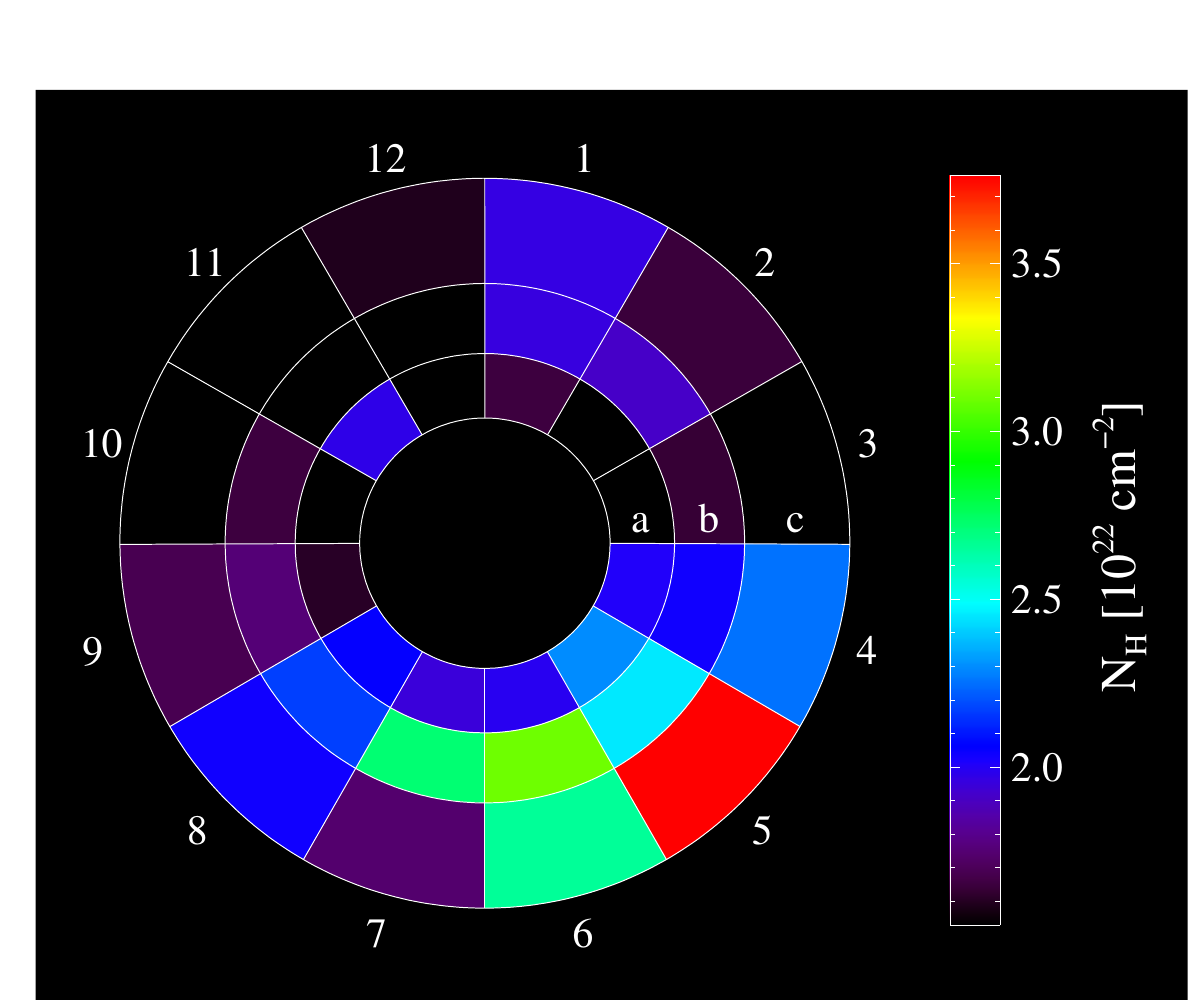}}
  \caption{Distribution of photo-electric absorption column $N_{\rm
      H}$ derived from the spectral fits shown in
    Fig.~\ref{fig:norms}. A clear excess of absorption is found in the
    lower half of the image, with a peak at position
    [c5]. \vspace*{6pt}}
  \label{fig:columns}
\end{figure}

By continuously scanning the telescope while making the observations
(on-the-fly mapping) large areas of sky can be readily mapped by
Mopra.  Data cubes covering $1^{\circ}\times 1^{\circ}$ of sky
typically take 4 nights with the telescope to obtain, and achieve a 1
sigma sensitivity of about 1.3K and 0.5K per $0.1\,{\rm km\,s^{-1}}$
velocity channel for the 12CO and the (13CO, C18O) lines,
respectively.  Such a data cube, covering the G322 region
($l=322^{\circ}-323^{\circ}$, $b=-0.5^{\circ}$ to $+0.5^{\circ}$), was
used for the analysis presented here.

\subsection{21 cm Data}

We use publicly available 21 cm radio data from the Parkes/ATCA
Southern Galactic Plane survey \citep{mcclure-griffiths:05} to
estimate the column density of neutral atomic hydrogen towards
Circinus X-1.  We use CO data to identify velocity components because
the velocity dispersion of the HI velocity components is too large to
unambiguously resolve individual clouds.

\section{Analysis}
\label{sec:analysis}
\subsection{X-ray Spectral Fits}
\label{sec:xray_spectra}
We performed spectral fits of data from {\em Chandra} ObsID 15801 and
{\em XMM} ObsID 0729560501, with results from the spectral analysis
presented in Figs.~\ref{fig:chandra_norms}-\ref{fig:columns}.  Because
the {\em XMM-Newton} FOV is contiguous and covers the rings entirely,
we extracted spectra for each ring of XMM ObsID 0729560501 in annuli
spanning the radial ranges [4.7'-7.1',7.1'-9.7',9.7'-13.7'], with each
ring divided into twelve sections.  Rings are labeled [a] through [c]
(with {\em lower case} letters denoting {\em XMM} rings), and the
layout of the grid is shown in Fig.~\ref{fig:norms}.  We divided the
{\em Chandra} FOV into a rectangular grid, with each CCD covered by an
8x8 grid of square apertures.  The layout of the grid on the ACIS
focal plane is shown in Fig.~\ref{fig:chandra_norms}, with columns
running from [A] through [P] and rows from 1 to 24 (with {\em capital}
letters denoting {\em Chandra} columns).

Spectra for {\em Chandra} ObsID 15801 were generated using the {\tt
  specextract} script after point source removal.  Background spectra
were generated from the blank sky background files.  No additional
background model was required in fitting the {\em Chandra} data, given
the accurate representation of the background in the black sky data.

Spectra for {\em XMM-Newton} ObsID 0729560501 were extracted from the
{\em XMM}-MOS data using the {\tt ESAS} package and following the
Diffuse Analysis Cookbook.  In spectral fitting of the {\em XMM} data,
we modeled instrumental features (Al and Si) and solar-wind charge
exchange emission by Gaussians with line energies and intensities tied
across all spectra and line widths frozen below the spectral
resolution limit of {\em XMM}-MOS at $\sigma=1\,{\rm eV}$.  We modeled
the sky background emission as a sum of an absorbed {\tt APEC} model
and an absorbed powerlaw.  Intensities of both background components
were tied across all ring sections.

We modeled the dust scattering emission in each region as an absorbed
powerlaw, with floating absorption column and normalization, but with
powerlaw index tied across all regions, with the exception of the
tiles covering the central source and the supernova remnant in the
{\em Chandra} image, which span a square aperture from positon [J14]
to [M18]. Tying the powerlaw photon index $\Gamma$ across all spectra is
appropriate because (a) the scattering angle varies only moderately
across the rings, with
\begin{equation}
  \theta_{\rm sc}=\sqrt{\frac{2c\Delta t}{\left(1-x\right)x D}}
\end{equation}
and (b) for the scattering angles considered here ($\theta_{\rm sc} >
10'$), the energy dependence of the scattering cross section is
independent of $\theta_{\rm sc}$, with $d\sigma/d\Omega \propto
E^{-2}$ \citep[e.g.][]{draine:03}.  While the source spectrum might
have varied during the outburst, we are integrating over sufficiently
large apertures to average over possible spectral variations.

\begin{figure}[t]
  \center\resizebox{\columnwidth}{!}{\includegraphics{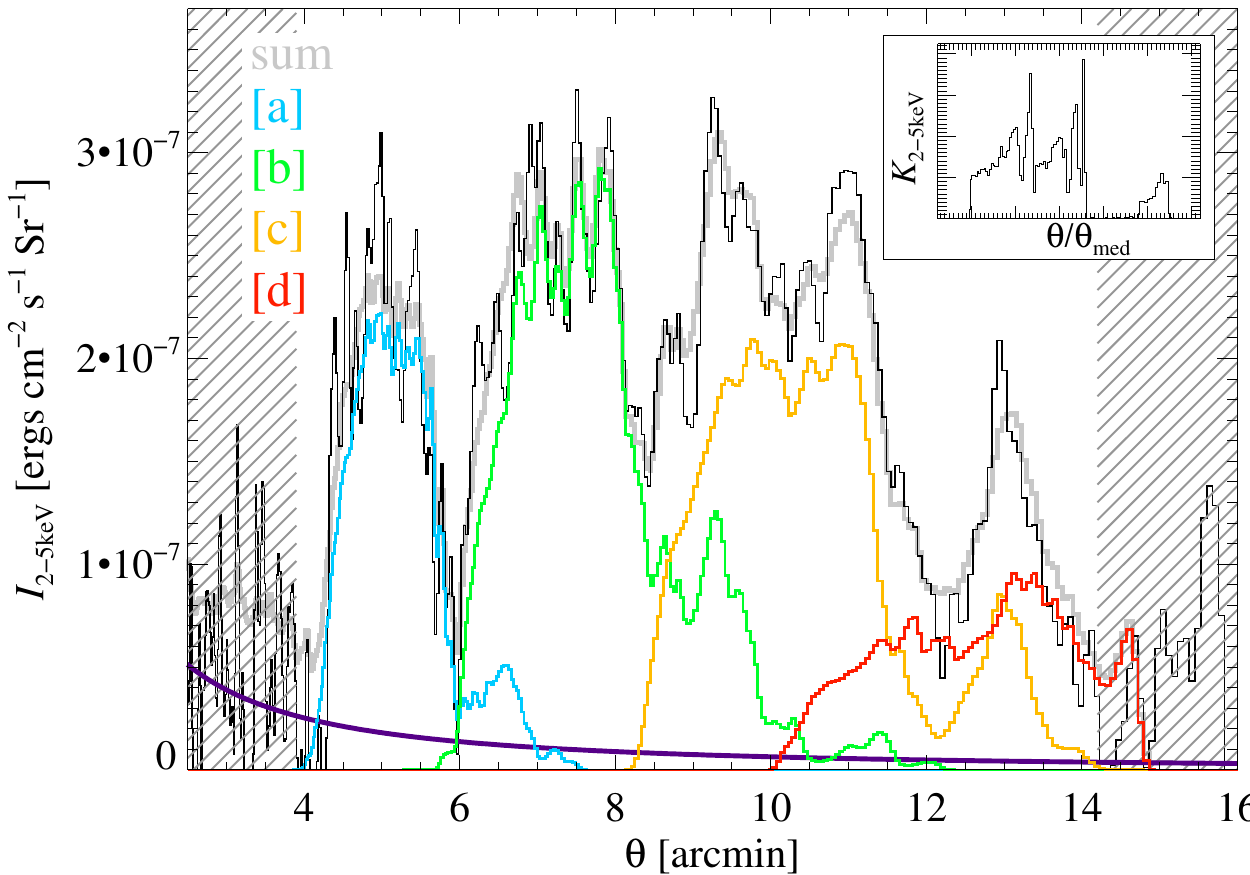}}
  \caption{Radial X-ray intensity profile of {\em Chandra} ObsID 15801
    in the 2-5keV band (chosen to roughly match the MAXI 2-4 keV
    band), shown in black. The intensity profile of {\em Chandra}
    ObsID 10062 was subtracted to remove the supernova remnant and
    residual quiescent background. Overlaid is the full reconstructed
    radial intensity profile from the deconvolution (gray), as well as
    the contributions from the four distinct dust concentrations along
    the line of sight identified in Fig.~\ref{fig:clouds} and the
    powerlaw model of the diffuse instantaneous dust scattering halo
    (purple line).  The hatched area shows the region where the
    deconvolution becomes unreliable because part of the kernel falls
    outside of the {\em Chandra} FOV or onto the supernova
    remnant. {\em Top right insert:} Dust scattering kernel
    $K(\theta/\theta_{\rm med})$ [see eq.~(\ref{eq:kernel})] used for
    the deconvolution, derived from the MAXI 2-4 keV lightcurve during
    the flare shown in Fig.~\ref{fig:maxi_lightcurve}, normalized by
    the total fluence of the outburst. \vspace*{6pt}}
  \label{fig:profile}
\end{figure}

While the source spectrum during outburst (the echo of which we are
observing in the rings) was likely more complicated than a simple
powerlaw, we have no direct measurement of the detailed flare spectrum
and the quality of the spectra does not warrant additional
complications, such as the addition of emission lines which are often
present in the spectra when the source is bright
\citep{brandt:00}. Dust scattering is expected to steepen the source
spectrum by $\Delta \Gamma =2$, and the measured powerlaw index of
\begin{equation}
  \Gamma = 4.00 \pm 0.03
\end{equation}
for the scattered emission is consistent with the typically soft
spectrum of $\Gamma \sim 2$ the source displays when in outburst.
 
The spectral fits resulting from the model are statistically
satisfactory, with a reduced chisquare of $\chi^{2}_{\rm
  red,0729560501} = 839.68/807\,{\rm d.o.f.} = 1.04$ for the combined
fit of the entire set of spectra for ObsID 0729560501 and
$\chi^{2}_{\rm red,15801} = 839.68/807\,{\rm d.o.f.} = 1.05$ for ObsID
15801.  We display the {\em unabsorbed\footnote{Unabsorbed intensities
    are calculated from the model fits by removing the effects of
    photo-electric absorption before determining model fluxes.}}
2-5keV intensity maps of {\em Chandra} ObsID 15801 and {\em XMM} ObsID
0729560501 in Figs.~\ref{fig:chandra_norms} and \ref{fig:norms},
respectively.  The unabsorbed intensity provides a direct measure of
the scattering intensity (unaffected by effects of foreground
absorption) and is directly proportional to the dust scattering depth.
In Figs.~\ref{fig:chandra_columns} and \ref{fig:columns}, we show maps
of the photo-electric absorption column density $N_{\rm H}$ determined
from the spectral fits for {\em Chandra} ObsID 15801 and {\em XMM}
ObsID 0729560501, respectively.

Four properties of the spectral maps stand out that confirm the
qualitative description of the color images in
Figs.~\ref{fig:color_chandra_image} and \ref{fig:color_xmm_image}:
\begin{enumerate}
\item{A clear intensity peak at position [a11] in Fig.~\ref{fig:norms}
    and [I19] in Fig.~\ref{fig:chandra_norms} corresponds to the
    brightness peaks seen at that position in
    Figs.~\ref{fig:color_chandra_image} and
    \ref{fig:color_xmm_image}. The intensity is a factor of two larger
    than in the neighboring section [a12] in Fig.~\ref{fig:norms},
    with a similar increment relative to the neighboring sections in
    Fig.~\ref{fig:chandra_norms}.}
\item{There is a clear intensity peak at position [b5] and [b6], with
    [b6] being approximately two times brighter than neighboring [b7]
    in Fig.~\ref{fig:norms}. Ring [b] is brighter on the Western side
    of the {\em XMM} FOV. This peak of ring [b] corresponds to the
    brightness peak in tiles [L9], [N10], and [O10] in
    Fig.~\ref{fig:chandra_norms}.}
\item{The Southern section of ring [c] is brighter than the Northern
    section, with the brightest scattering emission in sections
    [c5:c8] in Fig.~\ref{fig:norms}.  Ring [d] overlaps significantly
    with ring [c], and the color images and intensity maps cannot
    distinguish the angular distribution of the two rings.}
\item{There is a clear excess of foreground absorption in the Southern
    sections of the FOV, running from tile [P7] to [B6] in
    Fig.~\ref{fig:chandra_columns}. Figure \ref{fig:columns} shows the
    strongest absorption in sections [c5], [b6] and [b7]. This
    corresponds to the blue absorption lane that runs across the color
    images in Figs.~\ref{fig:color_chandra_image} and
    \ref{fig:color_xmm_image}.}
\end{enumerate}

The rings clearly deviate from axial symmetry about Circinus X-1,
which requires a strong variation in the scattering dust column
density across the image. Because the rings are well-defined, sharp
features, each ring requires the presence of a spatially well
separated, thin distribution of dust (where thin means that the extent
of the dust sheet responsible for each ring is significantly smaller
than the distance to the source, by at least an order of magnitude).
 
We present a quantitative analysis of the dust distribution derived
from the rings in \ref{sec:profiles}, but from simple consideration of
the images and spectral maps alone, it is clear that the different
dust clouds responsible for the rings have different spatial
distributions and cannot be uniform across the {\em XMM} and {\em
  Chandra} FOVs. In particular, the peak of the cloud that produced
the innermost ring (ring [a]) must have a strong column density peak
in section [a11], while the cloud producing ring [b] must show a
strong peak in sections [b5] and [b6]. The excess of photo-electric
absorption in the lower half of the image, with a lane of absorption
running through the sections [c5], [b6], and [b7], requires a
significant amount of {\em foreground} dust and gas (an excess column
of approximately $N_{\rm H}\sim 10^{22}\,{\rm cm^{-2}}$ relative to
the rest of the field).

\subsection{X-ray Intensity Profiles and Dust Distributions}
\label{sec:profiles}

\subsubsection{Radial intensity profiles}

In order to determine quantitatively the dust distribution along the
line of sight, we extracted radial intensity profiles of {\em Chandra}
ObsID 10062, 15801, and 16578 in 600 logarithmic radial bins from 1'
to 18'.5 in the energy bands 1-2 keV, 2-3 keV, and 3-5 keV (after
background subtraction and point-source/read-streak removal).  We
extracted the radial profile of {\em XMM} ObsID 0729560501 in 300
logarithmic bins over the same range in angles.  We subtracted the
quiescent radial intensity profile of ObsID 10062 to remove the
emission by the supernova remnant and residual background emission.

Because of the variation in photo-electric absorption across the
field, we restricted analysis to the 2-5 keV band, which is only
moderately affected by absorption. We plot the resulting radial
intensity profile of {\em Chandra} ObsID 15801 in
Fig.~\ref{fig:profile}. The plot in Fig.~\ref{fig:profile} shows four
echo rings labeled [a] through [d] which correspond to the visually
identified rings in the previous sections. Several bright sharp peaks
in $I_{2-5{\rm keV}}(\theta)$ are visible that can also be seen as
distinct sharp rings in Fig.~\ref{fig:color_chandra_image}, suggesting
very localized dust concentrations.  For comparison, the radial
profile of {\em XMM} ObsID 0729560501 is plotted in
Fig.~\ref{fig:xmm_profile}.

\subsubsection{Modeling and removing the diffuse instantaneous dust
  scattering halo}
In general, the dust scattering ring echo will be super-imposed on the
lower-level instantaneous diffuse dust scattering halo produced by the
(weak) emission of the point source during and just prior to the
observation, which should be removed before analysis of the echo. The
2-5 keV point source flux during and before ObsID 15801 was very low
($F_{2-5\,{\rm keV, 15801}} \lesssim 6\times 10^{-12}\,{\rm
  erg\,s^{-1}\,cm^{-2}}$),
resulting in essentially negligible halo emission, while the two later
observations showed larger point source flux
($F_{2-5\,{\rm keV}} \sim 1.8 \times 10^{-11}\,{\rm
  ergs\,s^{-1}\,cm^{-2}}$
for ObsID 16578 and
$F \sim 1.3\times 10^{-11}\,{\rm ergs\,s^{-1}\,cm^{-2}}$ for
0729560501, with correspondingly brighter diffuse halo emission).

We model the instantaneous dust scattering halo as a powerlaw in ring
angle, $I_{\rm halo} \propto \theta^{\eta}$, appropriate for the large
halo angles under consideration and consistent with the steady dust
scattering halo profile for Circinus X-1 derived from {\em Chandra}
ObsID 706 in \citet{heinz:13}.

We determine the powerlaw index to be $\eta \sim -1.5$ by fitting the
intensity profile of ObsID 0729560501 in the radial range between the
edge of the supernova remnant at $\theta > 2.5'$ and the inner edge of
ring [a] at $\theta < 5'$. This value of $\eta$ is consistent with the
reference dust scattering halo of ObsID 706 from \citet{heinz:13},
which has an index of $\eta \sim -1.55$ in the 2-5 keV band.

\begin{figure}[t]
  \center\resizebox{\columnwidth}{!}{\includegraphics{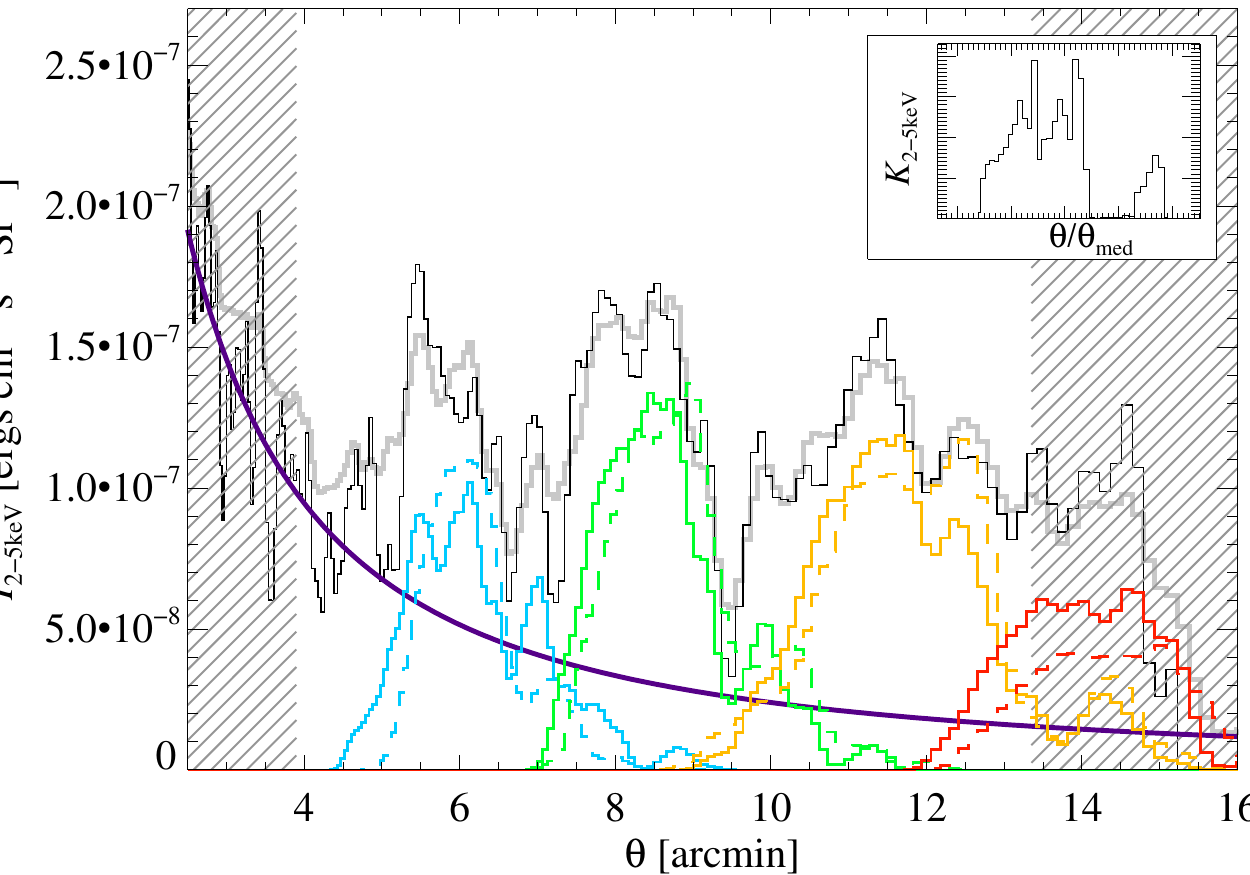}}
  \caption{ Radial X-ray intensity profile of {\em XMM} ObsID
    0729560501 in the 2-5keV band, following the same labeling and
    convention as Fig.~\ref{fig:profile}. The purple line shows the
    best fit powerlaw profile for the instantaneous dust scattering
    halo (removed before the deconvolution), the colored lines show
    the contributions from rings [a]-[d] derived from the radial
    deconvolution of the profile. Overplotted as dashed lines are the
    radial profiles from rings [a]-[d] predicted by the deconvolution
    of the {\em Chandra} ObsID 15801 radial profile. {\em Top right
      insert:} Dust scattering kernel $K(\theta/\theta_{\rm med})$
    [see eq.~(\ref{eq:kernel})] used for the deconvolution. \vspace*{6pt}}
  \label{fig:xmm_profile}
\end{figure}

We determine the powerlaw normalization for ObsID 15801 and 16578 by
fits to each radial profile between the outer edge of the remnant at
$\theta > 2.5'$ and the inner edge of ring [a] at $\theta < 4'$.  We
overplot the powerlaw fits to the instantaneous halo as purple lines
in Fig.~\ref{fig:profile} and \ref{fig:xmm_profile}. We subtract the
halo emission before further analysis of the ring echo in
\S\ref{sec:chandra_deconvolution} and \S\ref{sec:xmm_deconvolution}.

\subsubsection{Deriving the dust distribution}

Given the well-sampled MAXI lightcurve of the 2013 flare, we can
construct the light echo intensity profile that would be produced by a
thin sheet of dust at a given relative dust distance $x$ and observing
time $t_{\rm obs}$ from eqs.~(\ref{eq:theta}) and
(\ref{eq:intensity}). In general, the different parts of the flare
will generate emission at different radii (echoes of earlier emission
will appear at larger angles $\theta$).

The scattering angle and thus the scattering cross section will vary
for different parts of the lightcurve, which must be modeled.  Given
the large scattering angles of $\theta_{\rm sc} \gtrsim 1400''$ of the
rings and photon energies above 2 keV, it is safe to assume that the
cross section is in the large angle limit where
$d\sigma/d\Omega \propto \theta_{\rm sc}^{-\alpha} \propto
\theta^{-\alpha}$
with $\alpha\sim 4$ \citep{draine:03}, which we will use in the
following.  The exact value of $\alpha$ depends on the unknown grain
size distribution.  By repeating the analysis process with different
values of $\alpha$, ranging from 3 to 5, we have verified that our
results (in particular, the location of the dust clouds and the
distance to Circinus X-1) are not sensitive to the exact value of
$\alpha$.

For a thin scattering sheet at fractional distance $x$, the median
ring angle $\theta_{\rm med}$ (i.e., the observed angle $\theta$ for
the median time delay $\Delta t_{\rm med}$ of the flare) is
\begin{equation}
  \theta_{\rm med} = \sqrt{\frac{2c\Delta t_{\rm med}(1-x)}{xD}}
\end{equation}
where $\Delta t_{\rm med}$ is a constant set only by the date of the
observation and the date range of the flare.  The invariant intensity
profile $K_{2-5{\rm keV}}$ produced by that sheet depends
on the observed angle $\theta$ only through the dimensionless ratio $z
\equiv \theta/\theta_{\rm med}$:
\begin{eqnarray}
  \lefteqn{K_{2-5{\rm keV}}\left(z,x\right)} \nonumber \\
  \ \ \ & = & F_{\rm MAXI}(\Delta t_{\rm
    med} z^2)
  \frac{N_{\rm
      H,x}}{(1-x)^2}\left.\frac{d\sigma}{d\Omega}\right|_{\rm
    \theta=\theta_{\rm med}}z^{-4}
  \label{eq:kernel}
\end{eqnarray}

We can therefore decompose the observed radial profile into a series
of these contributions from scattering sheets along the line of sight,
each of the form $K_{2-5{\rm keV}}(\theta/\theta_{\rm
  med})$. Because we chose logarithmic radial bins for the intensity
profile, we can achieve this by a simple deconvolution of $I_{2-5{\rm
    keV}}$ with the kernel $K_{2-5{\rm keV}}$. The inset in
Fig.~\ref{fig:profile} shows $K(\theta/\theta_{\rm med})$ for
$\alpha=4$.

\subsubsection{Deconvolution of the {\em Chandra} radial profiles}
\label{sec:chandra_deconvolution}

To derive the dust distribution that gave rise to the light echo, we
performed a maximum likelihood deconvolution of the {\em Chandra}
radial profiles with this kernel into a set of 600 dust scattering
screens along the line of sight.  We used the Lucy-Richardson
deconvolution algorithm \citep{richardson:72,lucy:74} implemented in
the {\tt IDL ASTROLIB}
library\footnote{http://idlastro.gsfc.nasa.gov/ftp/pro/image/max\_likelihood.pro}. In
the region of interest (where the scattering profile is not affected
by the chip edge and the supernova remnant), the radial bins are
larger than the 50\% enclosed energy width of the {\em Chandra} point
spread function (PSF), such that the emission inside each radial bin
is resolved\footnote{We verified that our results are robust against
  PSF effects by repeating the deconvolution after smoothing both the
  profile and the kernel with Gaussians of equal width; this mimicks
  the effects PSF smearing would have on the deconvolution. The
  smoothing does not affect our results significantly even for widths
  an order of magnitude larger than the {\em Chandra} PSF.}.

Figure \ref{fig:profile} shows the reconstructed intensity profile
from this distribution for ObsID 15801 (derived by re-convolving the
deconvolution with $K_{\rm 2-5\,{\rm keV}}$ and adding it to the
powerlaw profile for the instantaneous dust scattering halo), which
reproduces the observed intensity profile very well, with only minor
deviations primarily for rings [a] from 4'-6', while the profile
outward of 6' is matched to within the statistical uncertainties.

\begin{figure}[t]
  \center\resizebox{\columnwidth}{!}{\includegraphics{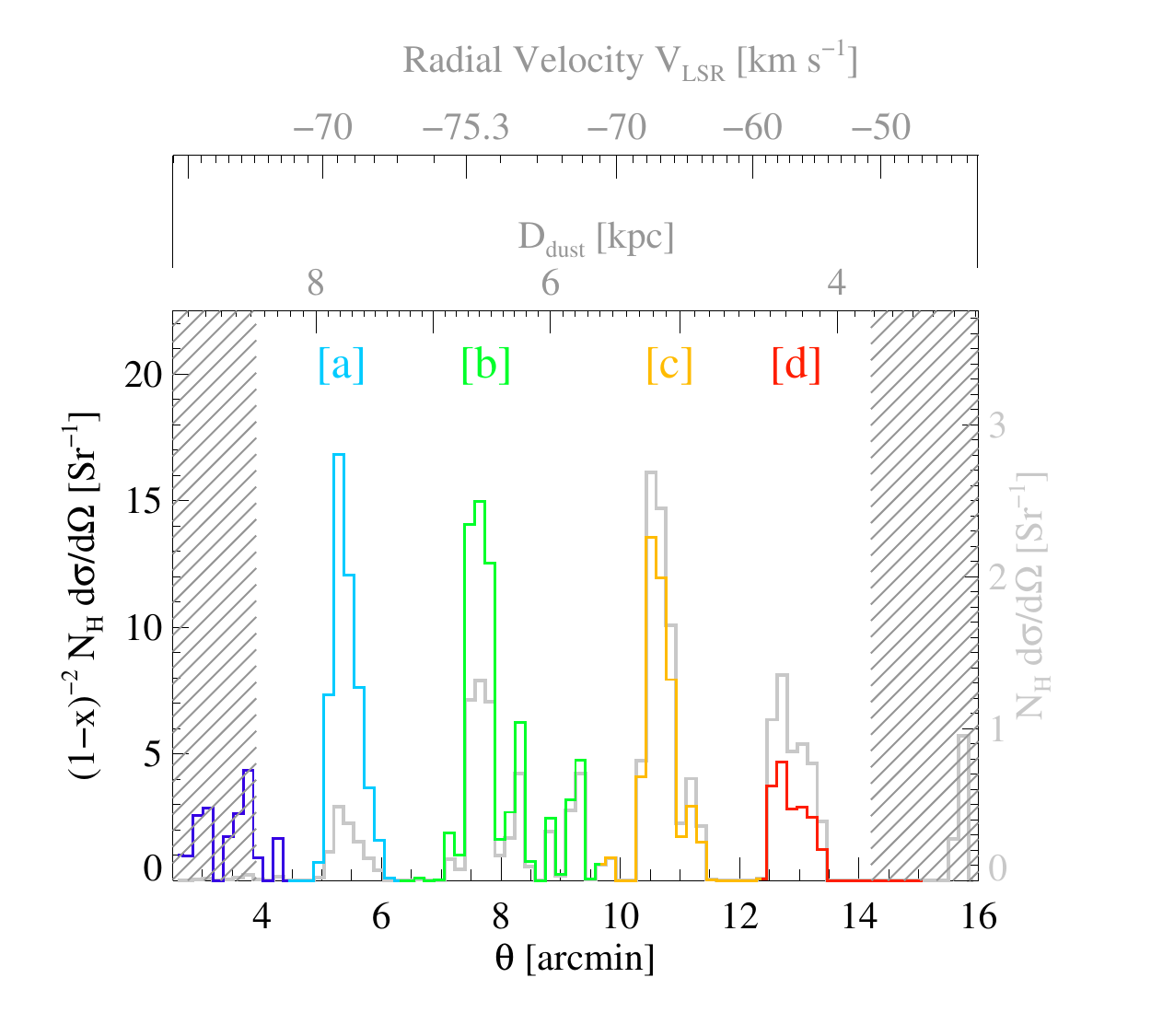}}
  \caption{Maximum likelihood deconvolution of the 2-5 keV radial
    intensity profile shown in Fig.~\ref{fig:profile}, plotted as a
    histogram of scattering depth along the line of sight (rebinned
    onto a uniform grid in $\theta$). Shown in color (left-hand
    y-axis) is the invariant effective scattering depth
    $(1-x)^{-2}N_{\rm H}d\sigma/d\Omega$, where $x$ is the fractional
    distance of the dust sheet relative to the distance to Circinus
    X-1 and $d\sigma/d\Omega$ is the scattering cross section
    evaluated at the mean scattering angle of the sheet and at 2
    keV. The distribution shows four main components (labeled [a]-[d]
    in colors blue, green, yellow, and red, respectively). Also shown
    in gray is the inferred absolute scattering depth $N_{\rm
      H}d\sigma/d\Omega$ for the best fit distance of $D_{\rm Cir
      X-1}=9.4{\rm kpc}$ (right-hand y-axis), along with the absolute
    distance to the scatterer and the associated radial velocity due
    to Galactic rotation (top axes in gray). \vspace*{6pt}}
  \label{fig:clouds}
\end{figure}

In Fig.~\ref{fig:clouds}, we plot the inferred dust distribution along
the line of sight as a function of median ring angle
$\theta_{\rm med}$, derived from our deconvolution of the intensity
profile of ObsID 15801, and rebinned onto a uniform grid in $\theta$.
The plotted quantity is a histogram of the invariant differential
scattering depth
$N_{\rm H}(1-x)^{2}\left.d\sigma/d\Omega\right|_{\theta=\theta_{\rm
    med}}$
per bin, which is independent of the distance to Circinus X-1.

\begin{deluxetable}{lcccc}
\tablecaption{Gaussian fits to dust
  clouds identified in the radial deconvolution in Fig.~\ref{fig:clouds}\label{tab:clouds}}

\tablehead{\colhead{Ring:} & 
  \colhead{$\theta_{\rm med}$} &
  \colhead{$\sigma_{\theta}$} &
  \colhead{$\frac{N_{\rm H}}{(1-x)^{2}}\frac{d\sigma}{d\Omega}
    [Sr^{-1}]$\tablenotemark{a}} &
  \colhead{\color{grey}{$x$}\tablenotemark{b}}}

\startdata \cutinhead{{\em Chandra} ObsID 15801}
{[}a] & $5.37' \pm 0.02'$ & $0.21' \pm 0.02'$ & $52.56 \pm 7.62$ & \color{grey}{$0.83$} \\
{[}b] & $7.63' \pm 0.01'$ & $0.19' \pm 0.01'$ & $47.86 \pm 4.77$ & \color{grey}{$0.70$} \\
{[}c] & $10.64' \pm 0.01'$ & $0.22' \pm 0.01'$ & $42.09 \pm 3.62$ & \color{grey}{$0.55$} \\
{[}d] & $12.85' \pm 0.04'$ & $0.34' \pm 0.04'$ & $19.60 \pm 3.17$ & \color{grey}{$0.45$} \\
\cutinhead{{\em Chandra} ObsID 16578}
{[}a] & $5.50' \pm 0.02'$ & $0.15' \pm 0.02'$ & $30.25 \pm 6.03$ & \color{grey}{$0.83$} \\
{[}b] & $7.91' \pm 0.02'$ & $0.20' \pm 0.02'$ & $36.16 \pm 4.18$ & \color{grey}{$0.70$} \\
{[}c] & $11.10' \pm 0.02'$ & $0.26' \pm 0.02'$ & $39.43 \pm 3.29$ & \color{grey}{$0.55$} \\
{[}d] & $13.49' \pm 0.02'$ & $0.10' \pm 0.02'$ & $6.51 \pm 1.68$ & \color{grey}{$0.45$} \\
\cutinhead{{\em XMM} ObsID 0729560501}
{[}a] & $6.00' \pm 0.05'$ & $0.90' \pm 0.06'$ & $29.31 \pm 2.32$ & \color{grey}{$0.84$} \\
{[}b] & $8.72' \pm 0.02'$ & $0.52' \pm 0.02'$ & $25.56 \pm 1.06$ & \color{grey}{$0.71$} \\
{[}c] & $12.04' \pm 0.02'$ & $0.73' \pm 0.02'$ & $23.54 \pm 0.92$ & \color{grey}{$0.56$} \\
\enddata
\tablecomments{Cloud centroids correspond to the median time delay of
  $\Delta t_{15801}=5.628\times 10^{6}\,{\rm s}$,
  $\Delta t_{16578}=6.203\times 10^{6}\,{\rm s}$, and
  $\Delta t_{0729560501}=7.925\times 10^{6}\,{\rm s}$ for ObsID 15801,
  16578, and 0729560501, respectively. We do not include ring [d] for
  {\em XMM} ObsID 0729560501 because its centroid falls outside of the
  MOS FOV.} \tablenotetext{a}{Invariant scattering depth, evaluated at
  the median scattering angle and at 2 keV energy.}
\tablenotetext{b}{relative distance $x$ to the dust cloud responsible
  for the respective ring using the best fit distance of
  $9.4\,{\rm kpc}$ [eq.~(\ref{eq:distance})]}
\end{deluxetable}

 The dust distribution clearly shows four main dust concentrations
toward Circinus X-1, which we plotted in different colors for clarity.
We fit each dust component with a Gaussian to determine the mean
location and width, with fit values listed in Table \ref{tab:clouds}.
Given the longer median time delay for ObsID 16578, the rings should
be observed at larger median angles by a factor of
\begin{equation}
      \label{eq:ring_increase}
      \frac{\theta_{\rm med,16578}}{\theta_{\rm med,15801}} =
      \sqrt{\frac{\Delta t_{\rm med,16578}}{\Delta t_{\rm med,15801}}} =
      \sqrt{\frac{6.084 \times 10^{6}\,{\rm s}}{5.628\times 10^{6}\,{\rm s}}}=
      1.04
    \end{equation}
    or 4\%, which is consistent with the observed increase in
    $\theta_{\rm ring,med}$, within the typical width of each cloud.

We label the identified dust clouds as components [a], [b], [c], and
[d] (plotted in green, blue, yellow, and red, respectively) according
to the naming convention adopted above, and identify them with the
main X-ray rings seen in the images. To show that this identification
is appropriate, we overplot the contribution of each dust cloud to the
total intensity profile in the associated color in
Fig.~\ref{fig:profile}. Rings [a] and [b] are produced primarily by
scattering from clouds [a] and [b], while ring [c] and [d] overlap
significantly and contain contributions from both clouds.

We tested the uniqueness of the deconvolution procedure using
synthetic radial profiles and verified that the procedure faithfully
reproduces the dust profile for randomly placed clouds down to a mean
flux of about 50\% of the background, for counts rates similar to the
ones observed in {\em Chandra} ObsID 15801.  Thus, while we cannot
rule out the presence of additional weak dust rings below the
detection threshold, the identification of the four main dust
concentrations producing rings [a]-[d] is robust.
    
{\subsubsection{Deconvolution of the {\em XMM} radial profile}
\label{sec:xmm_deconvolution}

The {\em XMM} data are noisier due to (a) the shorter exposure, (b)
the lower scattering intensity given the increased scattering angle at
the correspondingly longer time delay, (c) the larger background
intensity that is more difficult to model than the stable {\em
  Chandra} background, and (d) the brighter dust scattering halo due
to the increased point source flux during ObsID 0729560501.
    
Despite the lower spatial resolution and significantly lower
signal-to-noise ratio of the data, the rings are still discernible in
the radial profile in Fig.~\ref{fig:xmm_profile}.  We deconvolved the
residual intensity profile following the same procedure used for the
{\em Chandra} data While the 300 logarithmic radial bins used for
the deconvolution are narrower than the 50\% enclosed energy width of
{\em XMM}, we verified that the deconvolution was robust by confirming
that a strongly smoothed radial intensity profile and/or a smoothed
kernel lead to consistent locations and scattering depths within the
measurement uncertainties for the dust screens responsible for rings
[a]-[c].}.  The resulting contributions of rings [a] through [d] are
overplotted in color in Fig.~\ref{fig:xmm_profile}, following the same
convention as Fig.~\ref{fig:profile}.

Gaussian fits to rings [a], [b], and [c] are listed in Table
\ref{tab:clouds}. We do not include the fits for ring [d] because its
centroid falls too close to the edge of the FOV of {\rm XMM}-MOS at
$\theta \sim 14'.5$ (the {\em XMM}-PN FOV extends slightly further but
is too noisy by itself to reliably extend the profile outward).  The
inferred relative dust distances agree with the {\em Chandra} values
to better than 5\%, indicating that the deconvolution is robust, and
consistent with an increase in median ring radius from {\em Chandra}
ObsID 15801 to {\em XMM} ObsID 0729560501 of $\sim 16$\%.
    
As an additional consistency check, we constructed the radial
intensity profile predicted by the deconvolution of the {\em Chandra}
ObsID 15801 radial profile for the increased time delay of {\em XMM}
ObsID 0729560501 (including a decrease in surface brightness by a
factor of 1.8 due to the $\sim 16\%$ larger scattering angle, assuming
$\alpha=4$).  The predicted profiles for all four rings are plotted as
dashed lines in Fig.~\ref{fig:xmm_profile}.  The profiles of rings
[a]-[c] agree well with the {\em XMM} deconvolution, given (a) the
different coverage of the FOV by the {\em Chandra} and {\em XMM}
instruments, which should result in moderate differences between the
two data sets even if they had been taken at the same time delay and
(b) the different angular resolutions and the responses of the
telescopes.

\subsection{CO Spectral and Image Analysis}
\label{sec:co_analysis}

The deconvolution into dust sheets from \S\ref{sec:profiles} shows
that the scattering column density must be concentrated in dense, well
localized clouds.  Gas with sufficiently high column density (of order
$10^{21}\,{\rm cm^{-2}}$) and volume number density (of order
$100\,{\rm cm^{-3}}$) to produce the observed brightness and
arcminute-scale variations in the echo can be expected to be largely
in the molecular phase \citep[e.g.][]{lee:14}.

\begin{figure}[t]
  \center\resizebox{\columnwidth}{!}{\includegraphics{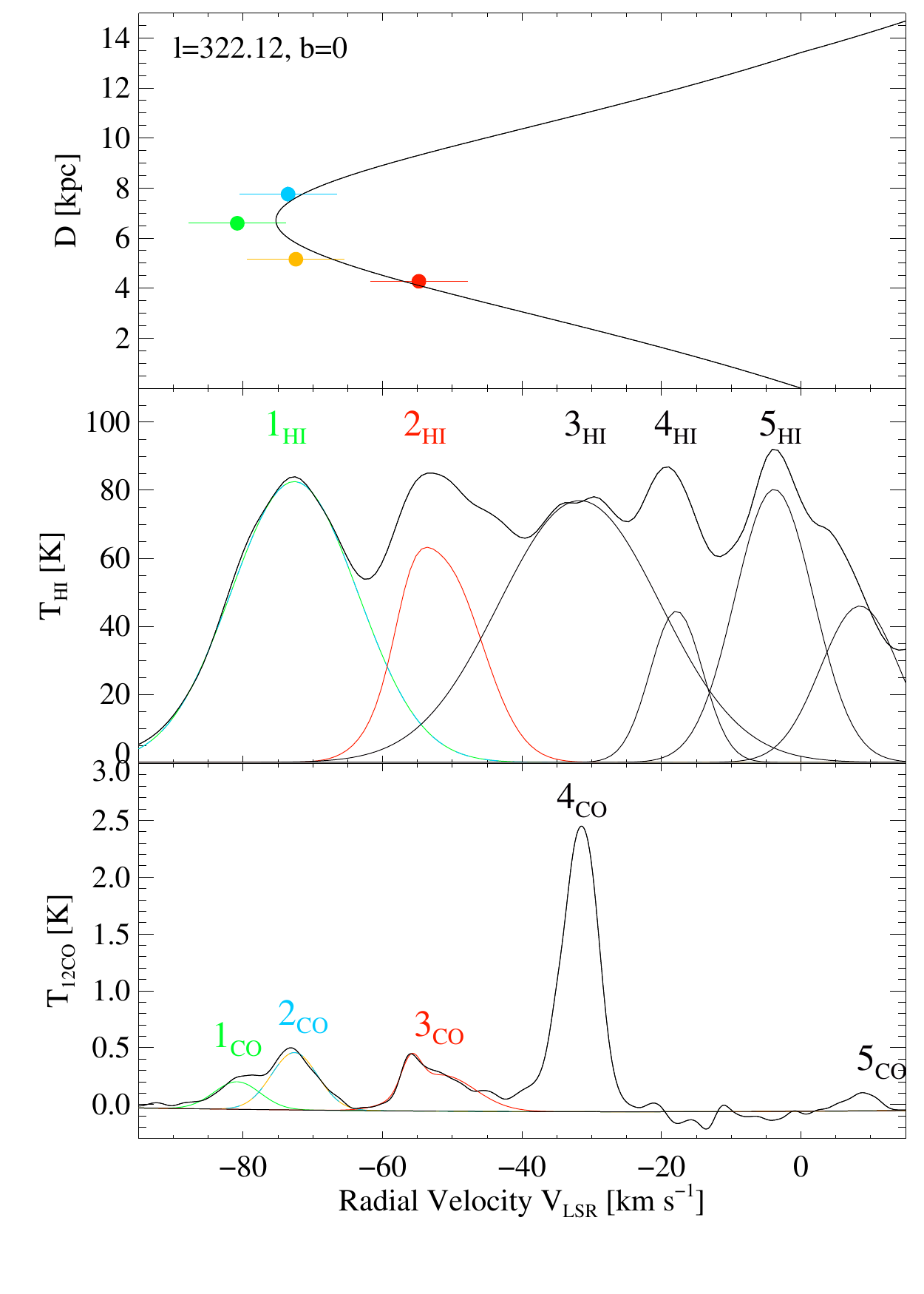}}
  \caption{{\em Top panel:} Radial velocity as a function of distance
    from the Sun for the sight-line in the direction of Circinus X-1
    at Galactic coordinates $l=322.12^{\circ},b=0^{\circ}$, using the
    rotation curves from \citet{mcclure:07} and \citet{brand:93} for
    the inner and outer Galaxy, respectively.  Overplotted are the
    locations and radial velocities for the four dust clouds [a]-[d]
    (from top to bottom) responsible for the dust scattering rings,
    assuming the best fit distance of 9.4 kpc. {\rm Middle panel:}
    21cm HI spectrum integrated over rings [a]-[c] defined in
    Fig.\ref{fig:norms}.  Component 1$_{\rm HI}$ comprises CO
    components 1$_{\rm CO}$ and 2$_{\rm CO}$ listed below. {\em Bottom
      panel:} $^{12}$CO spectrum integrated over the same region of
    rings [a]-[c] from Fig.~\ref{fig:norms}, smoothed with a $2\,{\rm
      km\,s^{-1}}$ FWHM Gaussian. Five discernible velocity components
    are labeled 1$_{\rm CO}$-5$_{\rm CO}$. \vspace*{6pt}}
  \label{fig:co_spectra}
\end{figure}

In order to search for corresponding concentrations of dust and
molecular gas, we compared the peaks in the inferred dust column
density to maps of CO emission from the Mopra Southern Galactic Plane
Survey that cover the entire {\em Chandra} and {\em XMM-Newton} FOV .

While CO emission does not directly probe the dust distribution along
the line of sight, the largest concentrations of dust will reside in
dense molecular gas, and CO is the most easily accessible tracer of
molecular gas that includes three-dimensional information about the
position of the gas, as well as a measure of the mass of the cold gas
through the CO intensity.

The bottom panel of Fig.~\ref{fig:co_spectra} shows the CO spectrum
integrated over the entire grid of rings identified in the {\em
  XMM-Newton} image in Fig.~\ref{fig:norms}, smoothed with a $2\,{\rm
  km\,s^{-1}}$ FWHM Gaussian for noise removal.  The spectrum shows
five well identified, localized components of molecular gas in the
direction of Circinus X-1.  A simple multi-Gaussian fit (shown in
color in the plot) locates the components at velocities $v_{\rm
  LOS,1}=-81\,{\rm km\,s^{-1}}$, $v_{\rm LOS,2}=-73\,{\rm
  km\,s^{-1}}$, $v_{\rm LOS,3}=-55\,{\rm km\,s^{-1}}$, $v_{\rm
  LOS,4}=-32\,{\rm km\,s^{-1}}$, and $v_{\rm LOS,5}=9\,{\rm
  km\,s^{-1}}$, where component 3$_{\rm CO}$ requires two Gaussian
components to fit at $-50\,{\rm km\,s^{-1}}$ and $-55\,{\rm
  km\,s^{-1}}$.

\begin{figure}[t]
  \center\resizebox{\columnwidth}{!}{\includegraphics{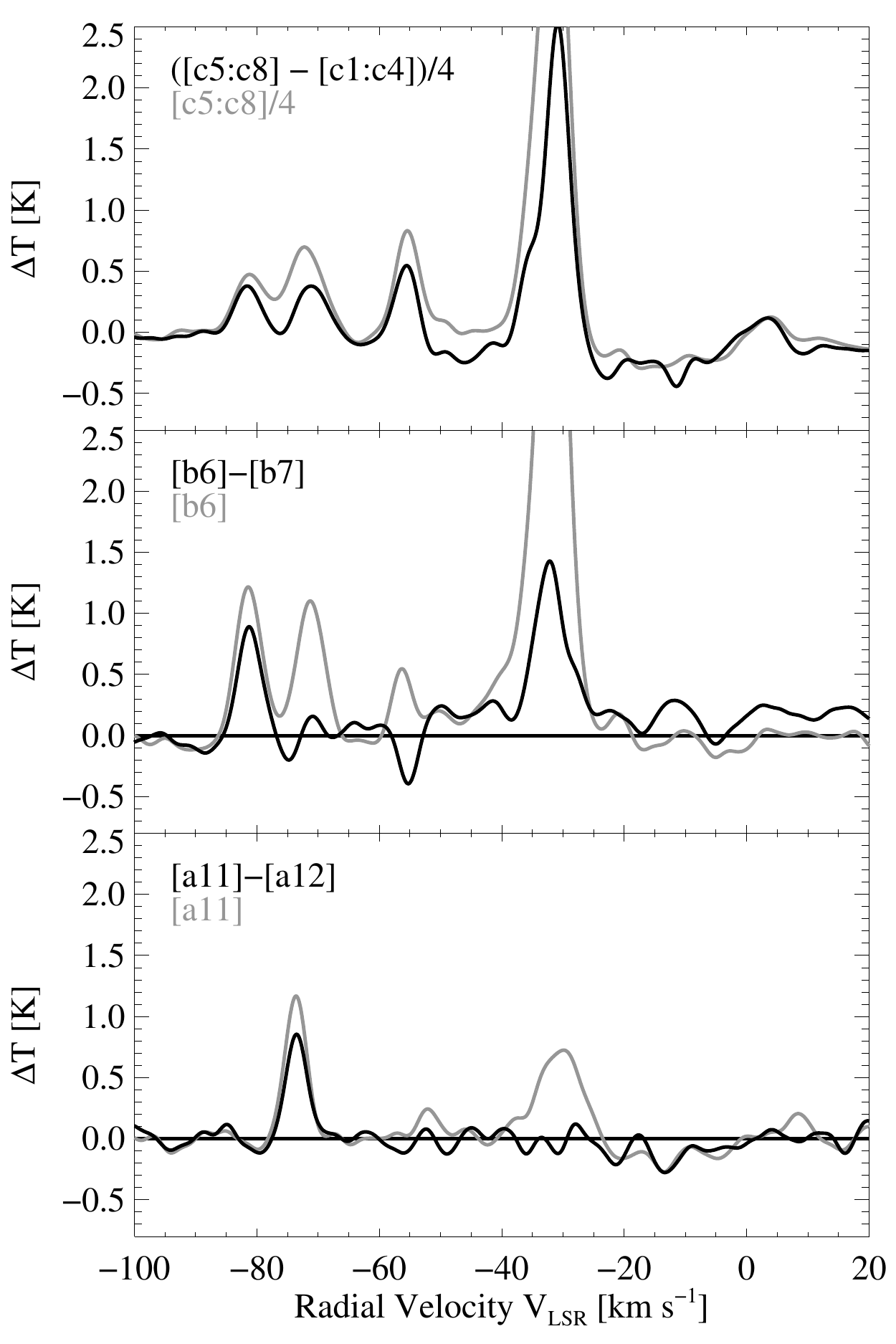}}
  \caption{Differential (black) and total (gray)  Mopra $^{12}$CO
    spectra of three ring sections (see Fig.~\ref{fig:norms} for a
    definition of the ring sections), smoothed with a $3\,{\rm
      km\,s^{-1}}$ FWHM Gaussian. Differential spectra show
    comparisons between neighboring ring sections that have large
    vs.~small dust scattering intensity to isolate the CO velocity
    component responsible for the X-ray peak. {\em Bottom:} ring
    section [a11] minus [a12], {\em middle:} ring section [b6] minus
    [b7]; {\em top:} sum of ring sections [c5:8] minus sum of
    [c1:4]. \vspace*{6pt}}
  \label{fig:differential_co_spectra}
\end{figure}
 
While other components of cold gas are likely present below the
sensitivity limit of the survey, and while some of the components may
be superpositions of several distinct clouds it is clear from the
spectrum that these are the primary concentrations of cold gas along
the line-of-sight and therefore the most probable locations of the
dust clouds responsible for the observed X-ray light echo.

We adopt the Galactic rotation curve by \citet{mcclure:07} for the
fourth Galactic quadrant within the Solar circle.  We use the rotation
curve by \citet{brand:93} for the outer Galaxy, matched at $v_{\rm
  LSR}=0$ following the procedure used in \citet{burton:13}, for the
Galactic position of Circinus X-1. The resulting radial velocity curve
in the direction of Circinus X-1 as a function of distance from the
Sun is plotted in the top panel of Fig.~\ref{fig:co_spectra}.  Because
the direction to Circinus X-1 crosses the Solar circle, the
distance-velocity curve is double-valued, and we cannot simply
associate a particular velocity with a given distance.

The rotation curve turns over at a minimum velocity of $v_{\rm rad}
\geq -75.3\,{\rm km\,s^{-1}}$.  Any velocity component at velocities
beyond the minimum {\em must} be displaced from the Galactic rotation
curve and lie near the tangent point. Given the velocity dispersion of
molecular clouds of $\sigma_{\rm CO} \lesssim 5\,{\rm km\,s^{-1}}$
relative to the LSR (see \S\ref{sec:discussion}), velocities of
$v_{\rm rad} \gtrsim -80\,{\rm km\,s^{-1}}$, as measured for component
1$_{\rm CO}$ are consistent with the rotation curve and place the
material at a distance of roughly $D\sim 6-8\,{\rm kpc}$.

\begin{figure}[t]
  \center\resizebox{\columnwidth}{!}{\includegraphics{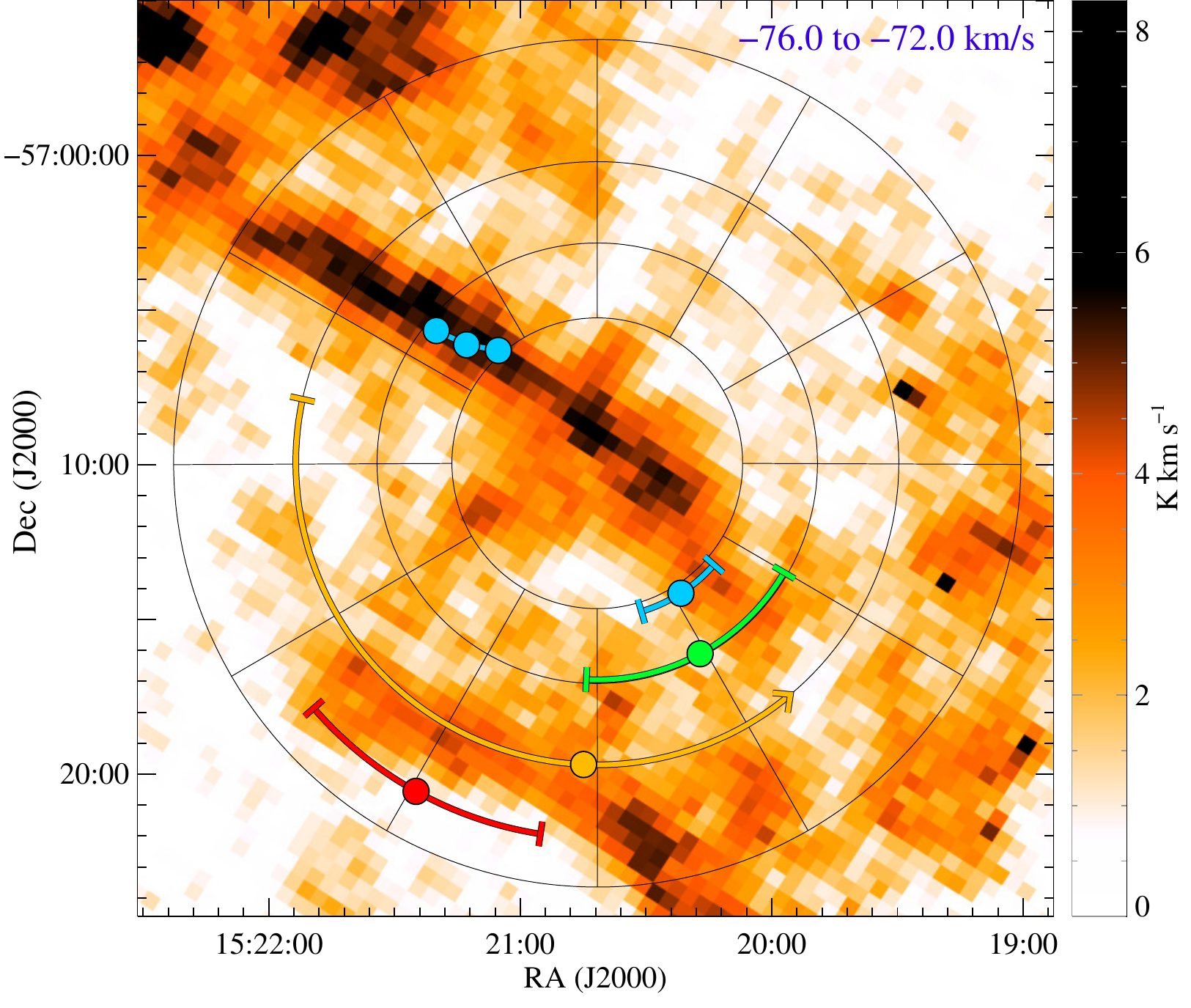}}
  \caption{Adaptively smoothed Mopra $^{12}$CO image of
    component 2$_{\rm CO}$ in the velocity range $-76\,{\rm
      km\,s^{-1}}$ to $-72\,{\rm km\,s^{-1}}$. The image shows a clear
    lane of dense molecular gas across the image. Overlaid for
    comparison is the angular grid used to denote rings in
    Fig.~\ref{fig:norms}. Connected blue dots are the local surface
    brightness peaks in ring [a11] from {\em Chandra} ObsID 15801,
    {\em XMM} ObsID 0729560501, and {\em XMM} ObsID 0729560601,
    showing a clear spatial coincidence with the CO lane. Also
    overlaid are the centroids (dots) and FWHMs (arcs) of the peaks in
    the angular intensity distributions of rings [a]-[d] from inside
    out in blue, green, yellow, and red, respectively, determined from
    spectral fits to {\em Chandra} ObsID 15801 in 10-degree segments
    for each ring. Note that the arc for ring [a] denotes the
    secondary peak. The peak of ring [c] is close to the chip edge,
    indicated by the arrow denoting the lower limit to the ring
    extent. \vspace*{6pt}}
  \label{fig:mopra_1}
\end{figure}

Given a rotation curve, one could attempt a global distance estimate
by simply comparing ring radii and CO velocities.  However, a
considerably more robust and accurate distance estimate can be derived
by also considering the angular distribution of dust inferred from the
brightness of each of the rings as a function of azimuthal angle.

\subsubsection{Differential CO spectra}
\label{sec:co_spectra}
\begin{figure}[t]
  \center\resizebox{\columnwidth}{!}{\includegraphics{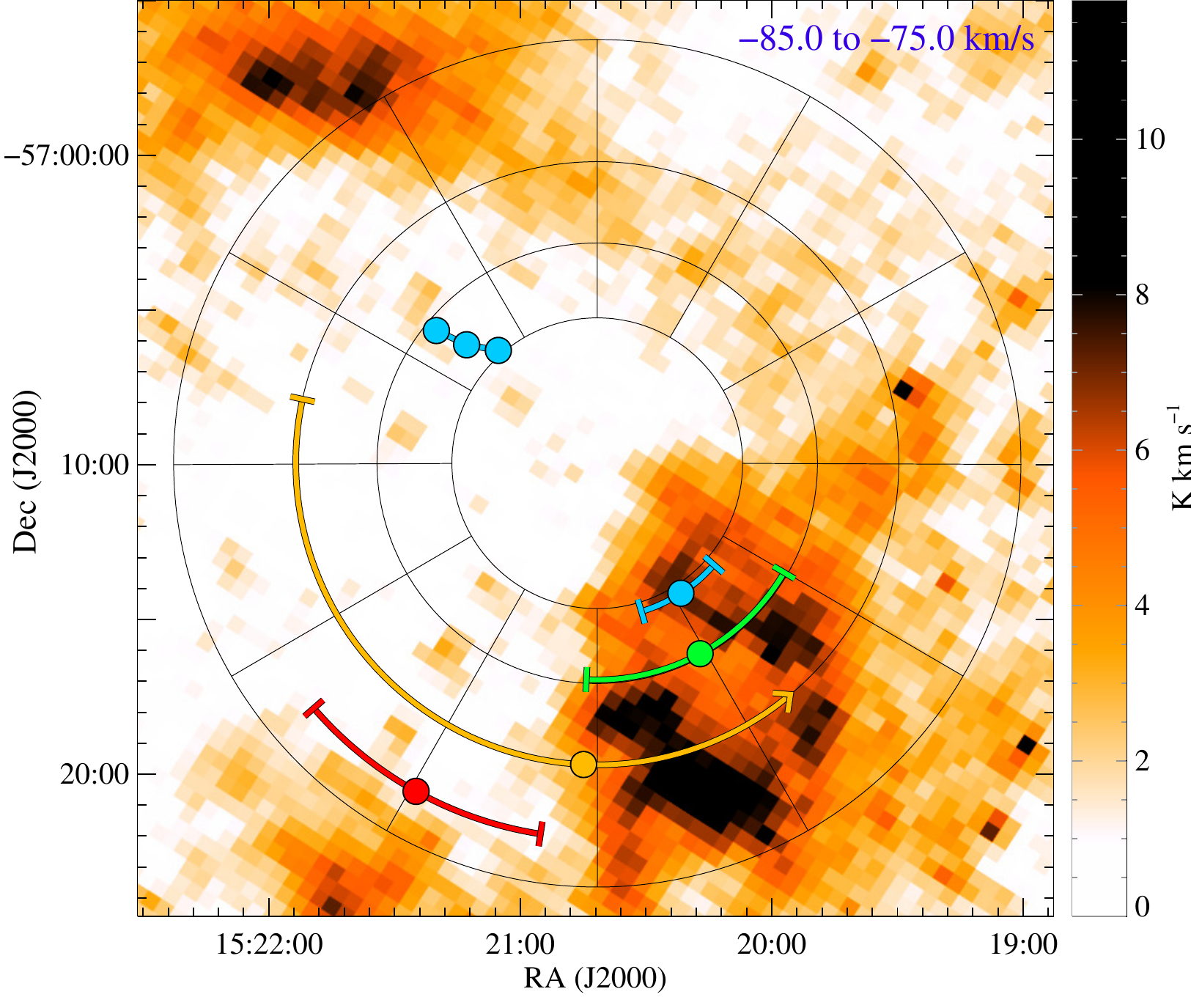}}
  \caption{Adaptively smoothed Mopra $^{12}$CO image of
    component 1$_{\rm CO}$ in the velocity range $-85\,{\rm
      km\,s^{-1}}$ to $-75\,{\rm km\,s^{-1}}$, same nomenclature as in
    Fig.~\ref{fig:mopra_1}. The CO emission in this band shows a clear
    peak at locations [b5] and [b6], matching the intensity
    distribution of X-ray ring [b] (green ring
    segment). \vspace*{6pt}}
  \label{fig:mopra_2}
\end{figure}

In order to identify which cloud is responsible for which ring, we
generated adaptively smoothed images of each velocity component within
velocity ranges centered on the CO velocity peaks of the clouds. The
images are shown in Figs.~\ref{fig:mopra_1}-\ref{fig:mopra_6} and
discussed in \S\ref{sec:co_images}. We also generated CO difference
spectra to identify which velocity components may be responsible for
the clear brightness enhancements seen in the different rings. They
are plotted in Fig.~\ref{fig:differential_co_spectra}, which we will
discuss below.

The un-absorbed X-ray intensity shown in Figs.~\ref{fig:chandra_norms}
and \ref{fig:norms} is directly proportional to the scattering dust
column density.  Variations in $I_{2-5keV}$ as a function of azimuthal
angle therefore probe the spatial distributions of the dust clouds
responsible for the different rings.  We can use a differential
$^{12}$CO spectrum of two ring sections that have significantly
different X-ray brightness to identify possible $^{12}$CO velocity
components that show a marked difference in column density consistent
with the dust enhancement.

\begin{deluxetable}{lccc}
\tablecaption{Gaussian fits to CO velocity components for rings
[a]-[d]\label{tab:co_velocities}}
 
\tablehead{ \colhead{CO cloud} &
\colhead{$v_{\rm CO} [{\rm km\,s^{-1}}]$} & 
\colhead{$\sigma [{\rm km\,s^{-1}}]$} & 
\colhead{$\frac{N_{\rm H_{2}}}{x_{\rm CO,20.3}} [10^{20}\,{\rm
cm^{-2}}]$\tablenotemark{a}}}

\startdata
\cutinhead{\bf Ring [a]:}
Cloud 2$_{\rm CO}$ & $-73.56 \pm 0.12$ & $1.46 \pm 0.12$ & $5.09 \pm 0.97$ \\
\cutinhead{\bf Ring [b]:}
Cloud 1$_{\rm CO}$ & $-80.85 \pm 0.19$ & $2.01 \pm 0.20$ & $3.88 \pm 0.97$ \\
\cutinhead{\bf Ring [c]:}
Cloud 1$_{\rm CO}$ & $-81.18 \pm 0.15$ & $1.72 \pm 0.16$ & $2.85 \pm 0.35$ \\
Cloud 2$_{\rm CO}$ & $-72.44 \pm 0.09$ & $2.35 \pm 0.09$ & $7.71 \pm 0.40$ \\
Cloud 3$_{\rm CO}$ & $-55.39 \pm 0.08$ & $1.82 \pm 0.08$ & $5.76 \pm 0.34$ \\
Cloud 4$_{\rm CO}$ & $-31.16 \pm 0.02$ & $2.16 \pm 0.02$ & $36.02 \pm 0.38$ \\
\cutinhead{\bf Ring [d]:}
Cloud 1$_{\rm CO}$ & $-81.29 \pm 0.16$ & $1.93 \pm 0.17$ & $3.43 \pm 0.76$ \\
Cloud 2$_{\rm CO}$ & $-72.14 \pm 0.11$ & $2.69 \pm 0.12$ & $7.59 \pm 0.90$ \\
Cloud 3$_{\rm CO}$ & $-54.80 \pm 0.09$ & $2.26 \pm 0.10$ & $7.10 \pm 0.78$ \\
Cloud 4$_{\rm CO}$ & $-30.85 \pm 0.02$ & $2.03 \pm 0.02$ & $34.22 \pm 0.73$  
\enddata
\tablecomments{$^{12}$CO spectra are integrated over rings of
        radii $4.0' < R_{a} < 6.0'$, $6.0' < R_{b} < 8.5'$, $8.5 <
        R_{c} < 12.0'$, and $11.0' < R_{d} < 14.0'$, respectively,
        weighted by the 3-5 keV {\rm Chandra} intensity.  Only
        velocity components identified in the differential CO spectra
        plotted in Fig.~\ref{fig:differential_co_spectra} as possible
        locations of dust responsible for the rings are listed}
      \tablenotetext{a}{hydrogen column
      density of the velocity component in units of $10^{20}\,{\rm
        cm^{-2}}/x_{\rm CO,20.3}$, where $x_{CO,20.3}$ is the CO to
      H$_{2}$ conversion factor in units of the fiducial value $x_{\rm
        CO} = 10^{20.3}\,{\rm cm^{-2}}/({\rm K\,km\,s^{-1}})$
      \citep{bolatto:13}.}
\end{deluxetable}

The two most obvious regions of strong local variation in
$I_{2-5\,{\rm keV}}$ are seen at ring sections [a11] and [b6], both of
which are significantly brighter than the neighboring sections [a12]
and [b7], respectively.

In the bottom panel of Fig.~\ref{fig:differential_co_spectra}, we show
the difference $^{12}$CO spectrum of rings sections [a11] minus [a12]
(i.e., the spectrum of [a12] subtracted from the spectrum of [a11]) as
a thick black line.  We also show the spectrum of [a11] only (i.e.,
without subtracting [a12]) as a thin gray line.  Because of the strong
enhancement in dust detected in [a11], the corresponding CO velocity
component must be significantly brighter in [a11] as well, and
therefore appear as a positive excess in the difference spectrum.  It
is clear from the figure that only {\em one} velocity component is
significantly enhanced in [a11] relative to [a12], the one at $v_{\rm
  rad}=-74\,{\rm km\,s^{-1}}$. We can therefore unequivocally identify
velocity component 2$_{\rm CO}$ at $-74\,{\rm km\,s^{-1}}$ as the one
responsible for ring [a].

The middle panel of Fig.~\ref{fig:differential_co_spectra} shows the
difference $^{12}$CO spectrum of [b6] minus [b7], where the light echo
is significantly brighter in [b6] than [b7]. In this case, two
$^{12}$CO velocity components are significantly enhanced in [b6]
relatively to [b7], component 1$_{\rm CO}$ at $v_{\rm rad}\sim
-80\,{\rm km\,s^{-1}}$ and component 4$_{\rm CO}$ at $v_{\rm rad}\sim
-32\,{\rm km\,s^{-1}}$. While we cannot uniquely identify which
velocity component is responsible for ring [b] from the spectra alone,
we can rule out components 2$_{\rm CO}$ and 3$_{\rm CO}$.

Finally, ring [c] does not show a single clearly X-ray enhanced
section relative to neighboring ones. Instead, the Southern sections
[c5:c8] are significantly enhanced relative to the Northern sections
[c1:c4]. We therefore plot the difference CO spectrum of sections [c5]
through [c8] minus sections [c1] through [c4] (top panel of
Fig.~\ref{fig:differential_co_spectra}). In this case, all four
velocity components are enhanced in [c5:c8] relative to [c1:c4], and
we cannot identify even a sub-set that is responsible for ring [c]
from the difference spectrum alone.

\subsubsection{CO images}
\label{sec:co_images}
In order to compare the ring emission with the spatial distribution of
cold gas and dust in the different velocity components identified in
Fig.~\ref{fig:co_spectra}, we extracted images around the peak of each
velocity component, adaptively smoothed to remove noise in velocity
channels of low-intensity while maintaining the full Mopra angular
resolution in bright velocity channels.  We employed a Gaussian
spatial smoothing kernel with width $\sigma=1.5' \times [1.0 - 0.9
\exp[-(v-v_{\rm peak})/(2\sigma_{\rm CO}^2)]$, where $v_{\rm peak}$
and $\sigma_{\rm CO}$ are the peak velocity and the dispersion of
Gaussian fits to the summed CO spectra shown in
Fig.~\ref{fig:differential_co_spectra}.  This prescription was chosen
heuristically to produce sharp yet low-noise images of the different
CO clouds.
 
\begin{figure}[t]
  \center\resizebox{\columnwidth}{!}{\includegraphics{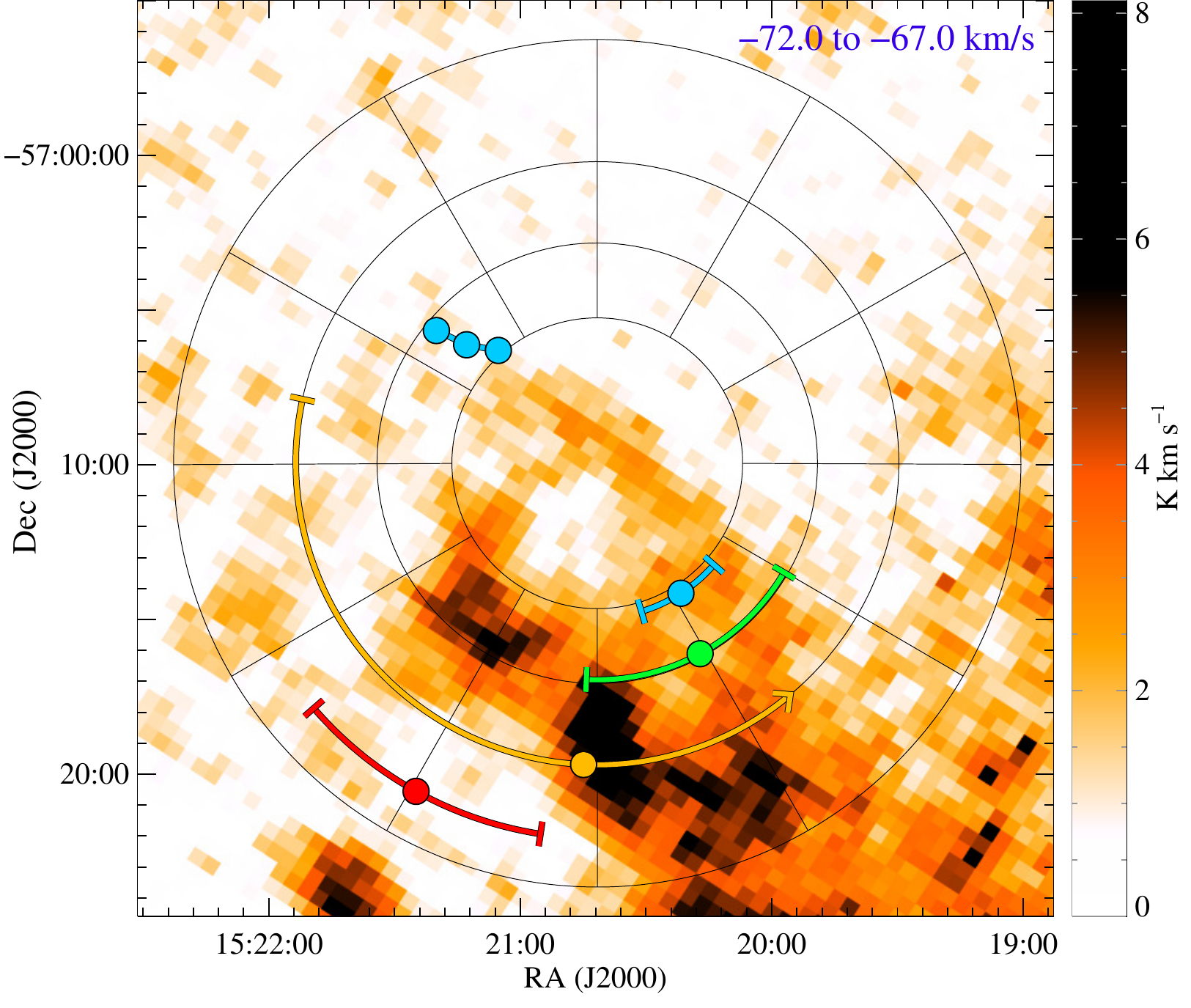}}
  \caption{Adaptively smoothed Mopra $^{12}$CO image of component
    2$_{\rm CO,b}$ in the velocity range $-72\,{\rm km\,s^{-1}}$ to
    $-67\,{\rm km\,s^{-1}}$, same nomenclature as in
    Fig.~\ref{fig:mopra_1}.  The CO intensity peak matches the
    centroid of X-ray ring [c] (yellow arc). \vspace*{9pt}}
  \label{fig:mopra_1b}
\end{figure}

\begin{figure}[t]
  \center\resizebox{\columnwidth}{!}{\includegraphics{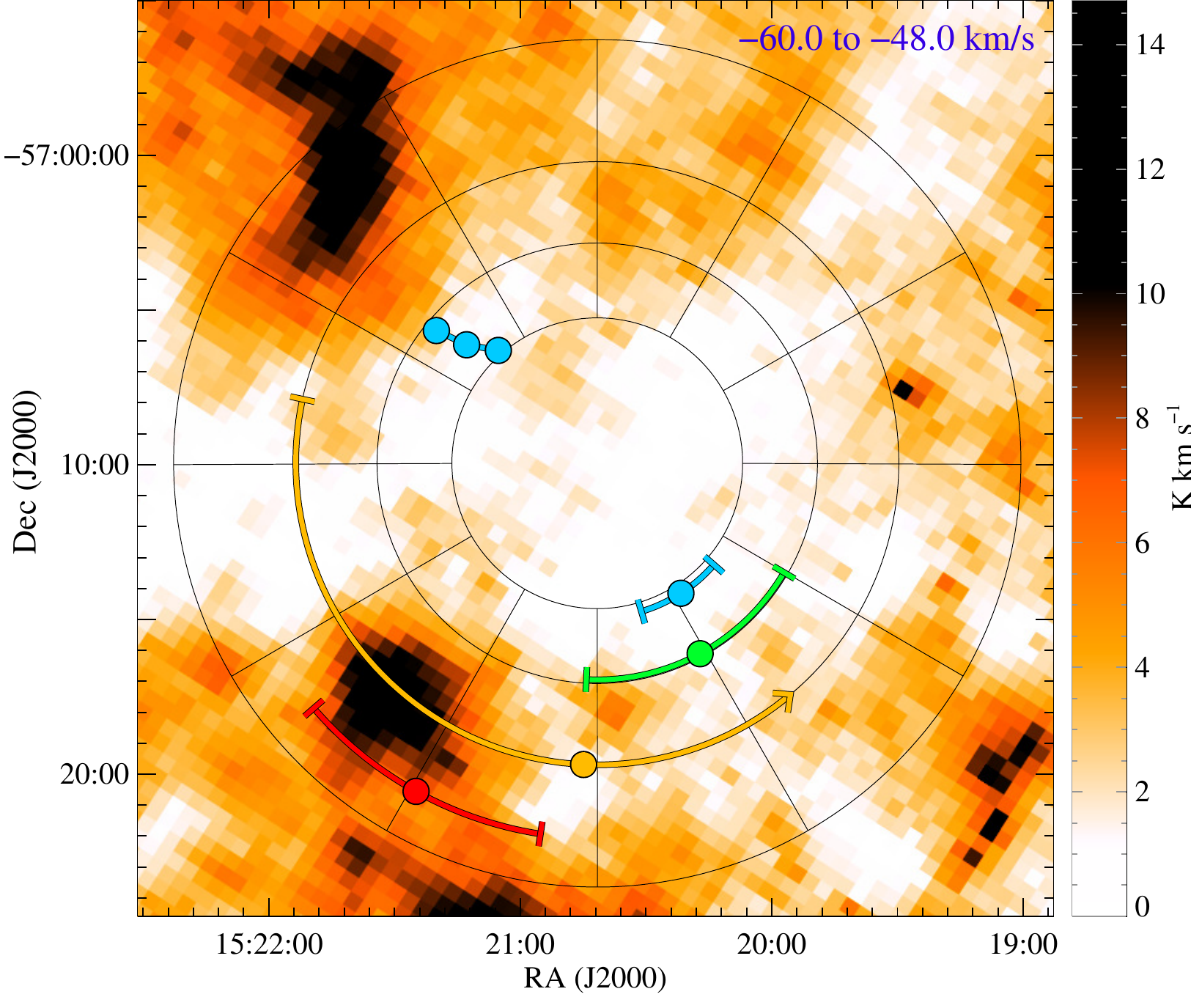}}
  \caption{Adaptively smoothed Mopra $^{12}$CO image of component
    3$_{\rm CO}$ in the velocity range $-60\,{\rm km\,s^{-1}}$ to
    $-48\,{\rm km\,s^{-1}}$, same nomenclature as in
    Fig.~\ref{fig:mopra_1}.  The CO intensity peak in the Southern
    half matches the location of the intensity peak of ring [d]
    (limited to the {\em Chandra} FOV; red arc). \vspace*{6pt}}
  \label{fig:mopra_3}
\end{figure}

\begin{figure}[t]
  \center\resizebox{\columnwidth}{!}{\includegraphics{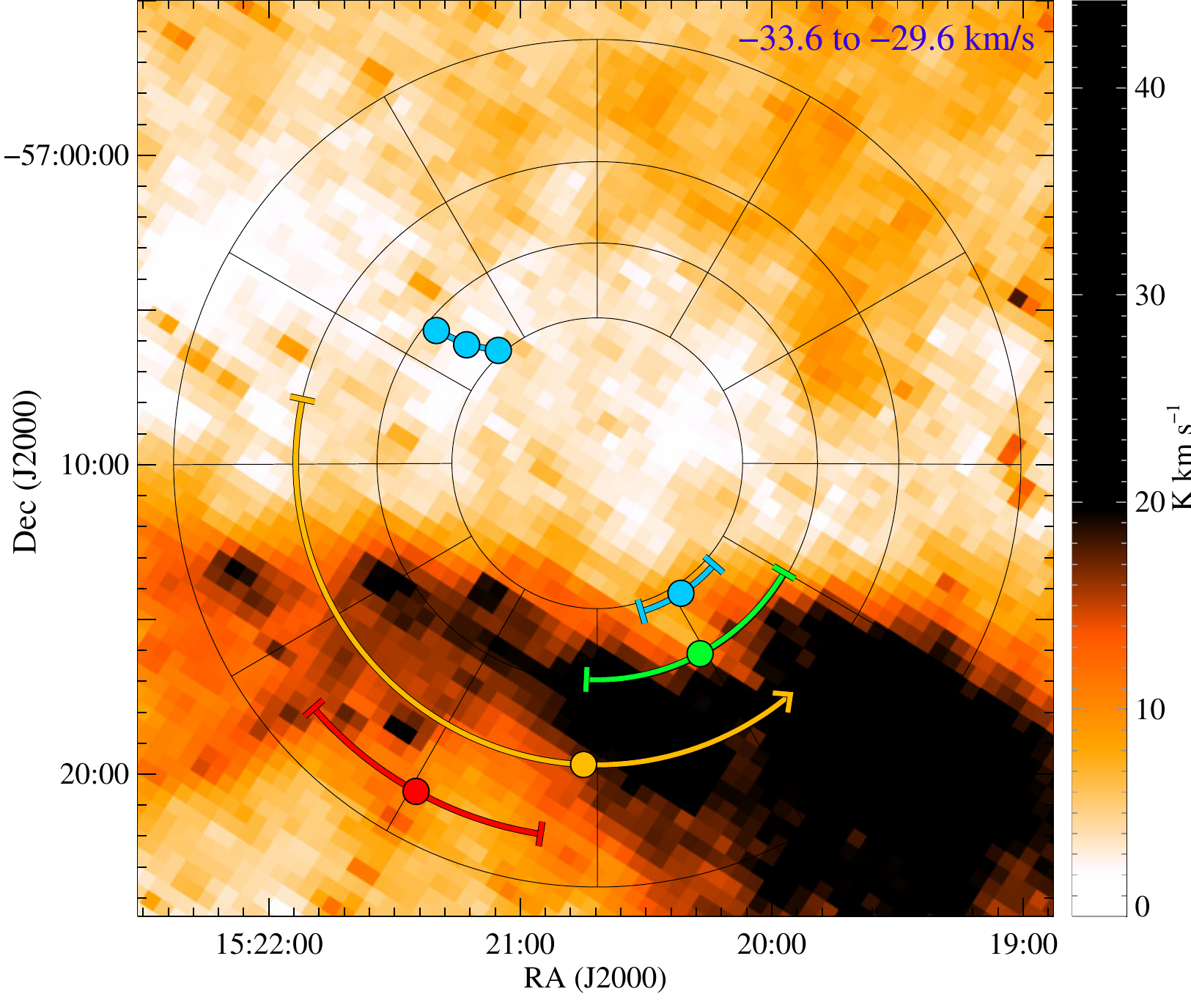}}
  \caption{Adaptively smoothed Mopra $^{12}$CO image of component
    4$_{\rm CO}$ in the velocity range $-33.6\,{\rm km\,s^{-1}}$ to
    $-29.6\,{\rm km\,s^{-1}}$, showing the brightest emission feature
    in the field of view; same nomenclature as in
    Fig.~\ref{fig:mopra_1}. The band of molecular gas across the lower
    half of the image matches the band of absorption seen in the color
    images in Fig.~\ref{fig:color_chandra_image} and
    \ref{fig:color_xmm_image} in position and morphology and
    corresponds to the excess absorption found in the spectrally
    determined absorption column shown in
    Figs.~\ref{fig:chandra_columns} and \ref{fig:columns}. \vspace*{6pt}}
  \label{fig:mopra_6}
\end{figure}

The CO intensity maps can be compared with the locations of excess
X-ray dust scattering to identify potential CO clouds responsible for
the different rings, supporting the spectral identification of
possible velocity components in \S\ref{sec:co_spectra} and
Fig.~\ref{fig:differential_co_spectra}.  To quantify the deviation of
the X-ray rings from axi-symmetry and for comparison with the CO data,
we constructed azimuthal intensity profiles of the rings from {\em
  Chandra} ObsID 15801 in ten degree bins, following the same spectral
fitting procedure used to construct Figs.~\ref{fig:chandra_norms} and
\ref{fig:chandra_columns}.  Ring radii in the ranges [a]: 4'-6', [b]:
6'-8', [c]: 9'-10'.5, [d]: 11'.5-12'.75 were chosen to best isolate
each ring.  The centroids and FWHM of the intensity peaks determined
from Gaussian fits for each ring are plotted as colored circles and
arcs in Figs.~\ref{fig:mopra_1} -\ref{fig:mopra_6}, respectively.

It is important to note that a one-to-one match in the intensity
distribution should not be expected on large scales for every ring for
three reasons: (a) multiple distinct clouds at {\em different
  distances} may fall into {\em the same} velocity channel because of
velocity deviations of the clouds from the local standard of rest; (b)
clouds at {\em different distances} may fall into the {\em same
  velocity channel} because of the double-valued nature of the
distance-velocity curve; (c) individual rings may contain scattering
contributions from multiple clouds at {\em different velocities} but
{\em similar distances}.

However, clear local maxima in scattering intensity may be expected to
correspond to local maxima in CO emission, and we will base possible
cloud-ring identification on local correspondence. Indeed, detailed
matches exist for rings [a] and [b], as already expected from the
spectra discussed in \S\ref{sec:co_spectra}, as well as for ring [d]:

\begin{itemize}
\item{Figure \ref{fig:mopra_1} shows an image of CO component
    2$_{\rm CO}$, integrated over the velocity range
    $-76\,{\rm km\,s^{-1}} < v_{\rm rad} < -72\,{\rm km\,s^{-1}}$,
    bracketing the velocity component at
    $v_{\rm rad} \sim -74\,{\rm km\,s^{-1}}$ identified in the
    differential spectra as giving rise to ring [a].  The clear
    association of ring [a] with component 2$_{\rm CO}$ at
    $-74\,{\rm km\,s^{-1}}$ in the CO image is striking. The prominent
    spectral feature at $v_{\rm rad} \sim -74\,{\rm km\,s^{-1}}$
    corresponds to a well-defined lane that runs through the position
    of \cir and crosses ring section [a11].  Overlaid as connected
    blue circles are the positions of the X-ray surface brightness
    peaks of ring [a] from {\em Chandra} ObsID 15801, and {\em XMM}
    ObsID 0729560501 and 0729560601 (from inside out, given the
    increasing ring radius at longer time delays for later
    observations)\footnote{{\em Chandra} ObsID 16578 is not shown
      because of the location of the peak at the chip gap and the
      increased noise at the chip boundary due to the shorter exposure
      time; {\em XMM} ObsID 0829560701 is not shown due to the
      overwhelming noise from the background flare that makes image
      analysis impossible.}. The intensity peak lies exactly on top of
    the CO peak and traces the CO lane as the rings sweep out larger
    radii.

    The obvious spatial coincidence and the fact that this is the {\em
      only} velocity component standing out in the differential
    spectrum of ring [a11] in Fig.~\ref{fig:differential_co_spectra}
    {\em unambiguously} determines that ring [a] must be produced by
    the dust associated with the CO cloud 2$_{\rm CO}$ at $v_{\rm rad}
    \sim -74\,{\rm km\,s^{-1}}$.

    In addition to the narrow lane in the North-Eastern quadrant of
    the image that we identify with the cloud responsible for ring
    [a], there is an additional concentration of CO emission in this
    channel in the Southern half of the image, which we identify as
    the cloud likely responsible for at least part of ring [c], and
    which we discuss further in Fig.~\ref{fig:mopra_1b}.  

    This velocity channel straddles the tangent point at minimum
    velocity $-75.3\,{\rm km\,s^{-1}}$ and may contain clouds of the
    distance range from 5kpc to 8 kpc (accounting for random motions).
    It is therefore plausible that multiple distinct clouds may
    contribute to this image and it is reasonable to associate
    features in this image with both rings [a] and [c].}

\item{Figure \ref{fig:mopra_2} shows the $^{12}$CO image of cloud
    1$_{\rm CO}$ in the velocity range $-85\,{\rm km\,s^{-1}} < v_{\rm
      rad} < -75\,{\rm km\,s^{-1}}$, roughly centered on the peak at
    $-81\,{\rm km\,s^{-1}}$. The peak of the emission falls into
    sectors [b5:b6] and [c5:c6], while there is consistent excess CO
    emission in the Eastern part of ring [b]. This spatial coincidence
    with the observed excess scattering emission of ring [b] strongly
    suggests that cloud 1$_{\rm CO}$ is responsible for the bulk of
    the X-ray scattering for ring [b], consistent with the difference
    spectrum in the middle panel of
    Fig.~\ref{fig:differential_co_spectra}.  Because of the strong
    excess foreground absorption in these ring sections (see
    Fig.~\ref{fig:columns}), a direct comparison with the intensity
    peaks in the X-ray images is more difficult than in the case of
    ring [a] (because the 3-5 keV channel images are significantly
    more noisy than the 1-2 keV and 2-3 keV channels, which are more
    strongly affected by absorption).  However the clear X-ray
    intensity peak in sections [b5:b6] denoted by the green arc
    determined from the spectra correlates well with the peak in CO
    emission of cloud 1$_{\rm CO}$.}

\item{Figure \ref{fig:mopra_1b} shows a low-band image of component
    2$_{\rm CO}$ in the velocity range $-72\,{\rm km\,s^{-1}} < v_{\rm
      rad} < -67\,{\rm km\,s^{-1}}$.  The CO emission peaks in
    sections [c5:c7] and corresponds to the centroid in the X-ray
    intensity of ring [c], which is overplotted as a yellow arc. The
    good spatial correspondence suggests that ring [c] is produced by
    a CO cloud at similar velocity but smaller relative distance than
    the CO cloud responsible for ring [a] shown in
    Fig.~\ref{fig:mopra_1}. Because of the substantial spectral
    overlap of both clouds, this image may not show the entire extent
    of the Southern part of component 2$_{\rm CO}$.}

\item{Figure \ref{fig:mopra_3} shows an image of cloud 3$_{\rm CO}$ in
    the velocity range $-60\,{\rm km\,s^{-1}} < v_{\rm rad} <
    -48\,{\rm km\,s^{-1}}$. The CO emission peaks in section [c8],
    coincident with the peak in the X-ray intensity of ring [d] at the
    same location (Fig.~\ref{fig:norms}).  

    Given the presence of several CO components in the Southern half
    of the image, identification of the intensity peaks of overlapping
    rings [c] and [d] is not as straight forward.  However, it is
    clear that not all of rings [c] and [d] can be explained by
    component 3$_{\rm CO}$ {\em alone}. An additional contribution
    from material in {\em other} velocity channels is required to
    explain the excess emission in section [c5:c6].  Our discussion of
    Fig.~\ref{fig:mopra_1b} suggests that the Southern component of
    1$_{\rm CO}$ is the second component responsible for rings [c]
    and/or [d].}

\item{Figure \ref{fig:mopra_6} shows an image of cloud 4$_{\rm CO}$ in
    the velocity range
    $-33.6\,{\rm km\,s^{-1}} < v_{\rm rad} < -29.6\,{\rm km\,s^{-1}}$,
    which is the dominant CO component in the Mopra FOV.  A very
    bright lane of molecular gas is crossing the Southern half of the
    FOV through sections [c5], [b5:b7].  Comparison with the X-ray
    images shows a clear spatial overlap of this CO cloud with the
    bluest sections of the ring emission that we identified as
    foreground absorption above. A contour of this cloud is overlaid
    on Figs.~\ref{fig:color_xmm_image} and \ref{fig:chandra_columns}
    and traces the region of highest column density very closely.

    Because component 4$_{\rm CO}$ causes absorption in rings [b],
    [c], and [d] without a correspondingly large scattering intensity
    peak (which would have to be about an order of magnitude brighter
    than the other rings, given the very large column density of
    component 4$_{\rm CO}$), the main component of cloud 4$_{\rm CO}$
    must be {\em in the foreground} relative to the clouds generating
    the scattering emission of X-ray rings [b], [c], and [d].  The
    brightest parts of this cloud are almost certainly optically thick
    in $^{12}$CO. The estimate of the cloud column density for this
    feature based on the $^{12}$CO intensity in Table
    \ref{tab:co_velocities} is therefore likely an underestimate. In
    fact, the differential hydrogen column inferred from the X-ray
    spectra, roughly $N_{\rm H} \sim 10^{22}\,{\rm cm^{-2}}$, is about
    a factor of two to three larger than that inferred from
    $^{12}$CO.}
\end{itemize}

From the image comparisons and the differential CO spectra, a clear
picture regarding the possible ring-cloud association emerges: Ring
[a] must be generated by cloud 2$_{\rm CO}$ at $\sim -74\,{\rm
  km\,s^{-1}}$. Ring [b] is generated by cloud 1$_{\rm CO}$ at $\sim
-81\,{\rm km\,s^{-1}}$.  At least part of the emission of ring [c] or
[d] is likely generated by cloud 3$_{\rm CO}$, but an additional
component is required to explain the emission in [c5:c6].

\section{Discussion}
\label{sec:discussion}

\subsection{The distance to \cir}
\label{sec:distance}
Based on the ring-cloud association discussed in the previous section,
we can use the kinematic distance estimate to each cloud from the
$^{12}$CO velocity and the relative distance $x$ to the cloud from the
light echo to constrain the distance to Circinus X-1.

To derive kinematic distances to the clouds, we assume that the
molecular clouds measured in our CO spectra generally follow Galactic
rotation, with some random dispersion around the local standard of
rest (LSR).  Literature estimates for the value of the 1D velocity
dispersion $\sigma_{\rm cloud}$ range from $\sigma_{\rm cloud} \sim
3.5\,{\rm km\,s^{-1}}$ \citep{liszt:81} for GMCs in the inner Galaxy
to $7\,{\rm km\,s^{-1}}$ \citep{stark:84,stark:89} for moderate-mass
high latitude clouds in the Solar Galactic neighborhood.  Recent
studies of kinematic distances to clouds assume a 1D dispersion of
$\sigma_{\rm cloud} \sim 3\,{\rm km\,s^{-1}}$ around the LSR
\citep[e.g.][]{roman-duval:09,foster:12}.  In the following, we will
conservatively adopt a value of $\sigma_{\rm cloud} \sim 5\,{\rm
  km\,s^{-1}}$.

In addition to random cloud motions, an unknown contribution from
streaming motions may be present.  The magnitude of streaming motions
relative to estimates of the Galactic rotation curve is generally
relatively poorly constrained observationally
\citep{roman-duval:09,reid:14}. \citet{mcclure:07} characterized the
possible contribution of streaming motions to deviations from the
rotation curve in the fourth quadrant and found systematic deviations
due to streaming motions with a peak-to-peak amplitude of $10\,{\rm
  km\,s^{-1}}$ and a standard deviation of $\sigma_{\rm streaming}
\sim 5\,{\rm km\,s^{-1}}$.  While streaming motions are not completely
random, the four dust clouds responsible for the rings are located at
sufficiently different distances that their streaming motions relative
to the Galactic rotation curve may be considered as
un-correlated. Thus, we will treat the uncertainty associated with
streaming motions as statistical and add the dispersion $\sigma_{\rm
  streaming}$ in quadrature to the cloud-to-cloud velocity dispersion,
giving an effective 1D bulk velocity dispersion relative to Galactic
rotation of $\sigma_{\rm cloud,eff}\sim 7\,{\rm km\,s^{-1}}$.

We assume that the main component of each dust cloud identified in
Fig.~\ref{fig:clouds} and Table \ref{tab:clouds} is associated with
one of the CO velocity components identified above.  However, a single
velocity component may be associated with {\em more} than one ring,
given the double-valued nature of the velocity curve and the possible
overlap of clouds in velocity space due to excursions from the LSR.
For a given a distance $D$ to Circinus X-1, we can then determine the
radial velocity of each dust cloud if it were at the LSR, given its
relative distance $x(D,\theta_{\rm med})$, and compare it to the
observed radial velocities of the CO clouds.

For an accurate comparison, we extracted CO spectra that cover the FOV
of {\em Chandra} ObsIDs 15801 and 16578, limited to the radial ranges
[a]: 4.0'-6.0', [b]: 6.0'-8.5', [c]: 8.5'-12', [d]: 11.0'-14.0' for ObsId
15801 and increase radial ranges by 4\% for ObsID 16578 according to
eq.~(\ref{eq:ring_increase}).  We weighted the CO emission across the
FOV with the 3-5~keV {\em Chandra} intensity in order to capture CO
emission from the regions of brightest dust scattering.  We fitted
each identified CO component with a Gaussian and list the relevant CO
clouds used in the fits in Table \ref{tab:co_velocities}.

For a given assumed distance $D$ to Circinus X-1, we calculate the
expected radial velocities of the cloud components and compare them to
the observed CO velocities to find the distance that best matches the
observed distribution of clouds.  We represent each CO cloud $j$ and
each dust cloud $i$ by the best-fit Gaussians from Tables
\ref{tab:co_velocities} and \ref{tab:clouds}, respectively, with
dispersions $\sigma_{\rm CO,j}$ and $\sigma_{\rm dust,i}$ and peak
positions $v_{\rm CO,j}$ and $v_{\rm dust,i}$, respectively.  In
addition, we allow for an effective bulk 1D velocity dispersion of
$\sigma_{\rm cloud,eff} = 7\,{\rm km\,s^{-1}}$ relative to the
Galactic rotation curve as discussed above, comprised of the random
cloud-to-cloud velocity dispersion and contributions from streaming
motions.

\begin{figure}[t]
  \center\resizebox{\columnwidth}{!}{\includegraphics{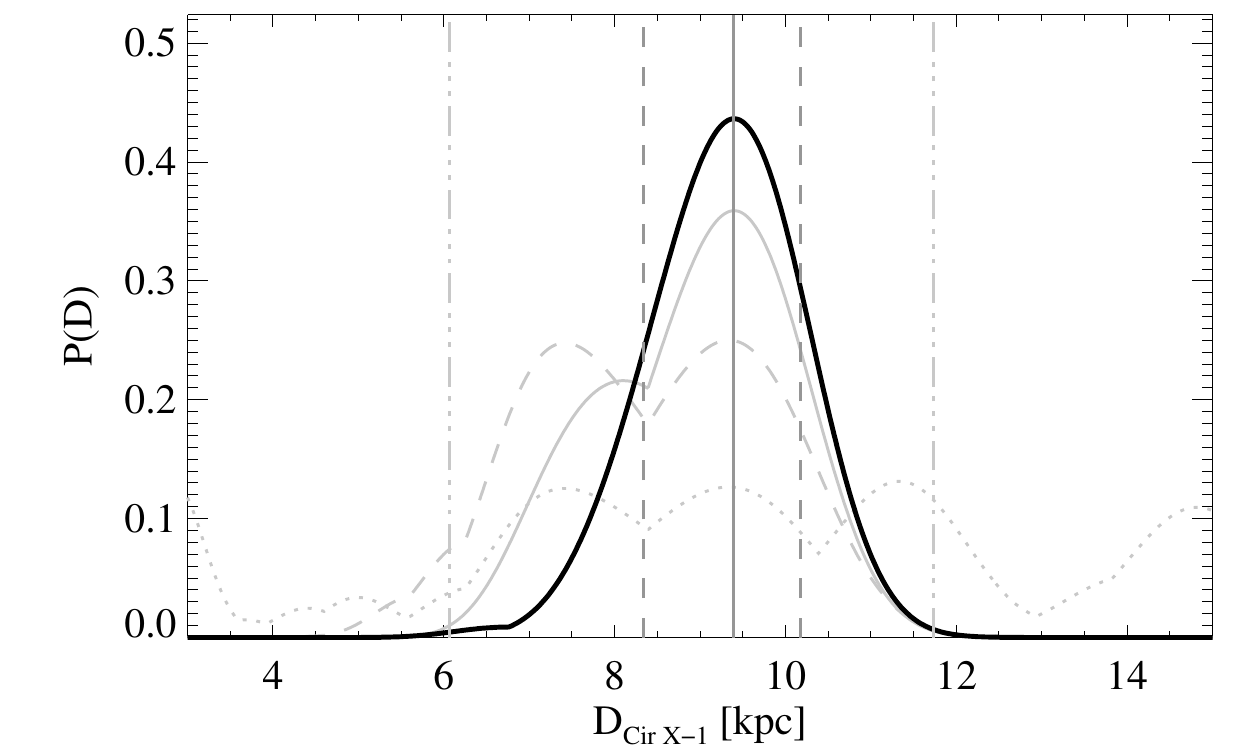}}
  \caption{Unnormalized posterior of the distance to Circinus
    X-1 from ObsID 15801 and 16578.  Shown in black: Likelihood
    function with all three priors about ring-CO-cloud associations in
    place (i.e., we identify velocity component 2$_{\rm CO}$ at
    $v_{\rm LOS}\approx -74\,{\rm km\,s^{-1}}$ with ring [a],
    component 1$_{\rm CO}$ at $v_{\rm LOS} \approx -80\,{\rm
      km\,s^{-1}}$ with the peak of ring [b], and require ring [c] and
    [d] to be produced by different CO components).  The likelihood
    distributions without {\em any} priors, with only prior {\em one}, and
    with prior {\em one and two} are plotted as dotted, dashed, and solid
    gray curves, respectively. The most likely distance of
    $9.4\,{\rm kpc}$ and the 1-sigma and 3-sigma uncertainties are plotted as
    solid, dashed, and dashed-triple-dotted gray vertical lines,
    respectively. \vspace*{6pt}
    \label{fig:likelihood}}
\end{figure}

While the velocity distribution of each dust cloud (calculated from
the Gaussian fit to the distribution of dust as a function of $z$) is
not itself Gaussian, we conservatively use the peak value of the
velocity and assign a velocity dispersion $\sigma_{\rm dist,i}$ that
is the larger of the two propagated $z$-dispersions in the positive
and negative $z$-direction from the peak of the cloud.

We combine the measurements on the ring radii from the deconvolutions
of ObsID 15801 and 16578, corrected by the expansion factor of the
rings given by eq.~(\ref{eq:ring_increase}). The goodness of fit is
then determined by the likelihood function $P(D)$ as a function of
distance $D$,
\begin{eqnarray}
  \label{eq:likelihood}
  \lefteqn{P(D)  =  \frac{1}{N_{P}} \times...} \\
& & ...\,{\Pi_{i=1}^{4} \,{\rm max}_{j_{i}=1}^{4}\left\{\exp{\left[-\frac{\left(v_{\rm
    dust,i}(D) - v_{\rm CO,j_{i}}\right)^2}{2\left(\sigma_{\rm
    dust,i}^2+\sigma_{\rm CO,j_{i}}^2 + \sigma_{\rm
    cloud}^2\right)}\right]}\right\}} \nonumber
\end{eqnarray}
where for each dust cloud $i$ in the product, the CO component $j_{i}$
with the maximum likelihood is chosen from the set of allowed CO
components, i.e., the CO component that most closely matches the
velocity $v_{\rm dist,i}$ of dust cloud $i$; $N_{P}$ is a
normalization factor chosen to ensure that $\int dD P(D) = 1$. $P(D)$
is plotted in Fig.~\ref{fig:likelihood}.

As discussed in \S\ref{sec:co_images}, we can restrict the CO clouds
responsible for the rings based on the visual and spectral
identification discussed in \S\ref{sec:co_spectra} and
\S\ref{sec:co_images}.  We place two strong priors and one weak prior
on the association of dust and CO clouds:
\begin{enumerate}
\item{The first strong prior is the association of ring [a] with CO
    component 2$_{\rm CO}$}
\item{The second strong prior is the association of ring [b] with CO
    component 1$_{\rm CO}$}
\item{The third, weak prior is the requirement that ring [c] and [d] cannot
    be produced by component 3$_{\rm CO}$ alone, i.e., that either
    ring [c], or ring [d] are associated with a component other than
    3$_{\rm CO}$.}
\end{enumerate}
Functionally, these priors take the form of limits on the possible
combinations of rings $i$ and clouds $j_{i}$ in the maximum taken in
eq.~(\ref{eq:likelihood}).  That is, the first prior eliminates the
maximum for ring $i=1$ and simply replaces it with the exponential for
cloud $j_{1}=2$, while the second prior replaces the maximum for ring
$i=2$ with the exponential for $j_{2}=1$. The third prior eliminates
combinations in the maximum for which $j=3$ for both $i=3$ and $i=4$.

\subsubsection{Distance estimate with all three priors}
\label{sec:all_priors}

With all three priors in place, eq.~(\ref{eq:likelihood}) takes the
form
\begin{eqnarray}
  \label{eq:likelihood_priors}
  \lefteqn{P(D)  =  \frac{1}{N_{P}}\times...}\\
& & ...\exp{\left[{-\frac{\left(v_{\rm
    dust,1}(D) - v_{\rm CO,2}\right)^2}{2\left(\sigma_{\rm
    dust,1}^2+\sigma_{\rm CO,2}^2 + \sigma_{\rm
    cloud,eff}^2\right)}}\right]} \times...\nonumber \\
& & ...\exp{\left[{-\frac{\left(v_{\rm
    dust,2}(D) - v_{\rm CO,1}\right)^2}{2\left(\sigma_{\rm
    dust,2}^2+\sigma_{\rm CO,1}^2 + \sigma_{\rm
    cloud,eff}^2\right)}}\right]}\times ...\nonumber \\
& & ... \Pi_{i=3}^{4}\,{\rm max}_{j_{i}=1}^{4}\left\{\left(1 -
    \delta_{j_{3},3}
    \delta_{j_{4},3}\right)\times ... \vphantom{\exp{\left[-\frac{\left(v_{\rm
    dust,i}(D) - v_{\rm CO,j_{i}}\right)^2}{2\left(\sigma_{\rm
    dust,i}^2+\sigma_{\rm CO,j_{i}}^2 + \sigma_{\rm
    cloud,eff}^2\right)}\right]}}\right.\nonumber \\ 
  & & \ \ \ \ \ \ \ \ \ \ \ \ \ \ \ \left. ... \exp{\left[-\frac{\left(v_{\rm
    dust,i}(D) - v_{\rm CO,j_{i}}\right)^2}{2\left(\sigma_{\rm
    dust,i}^2+\sigma_{\rm CO,j_{i}}^2 + \sigma_{\rm
    cloud,eff}^2\right)}\right]}\right\} \nonumber
\end{eqnarray}
where the product $\delta_{j_{3},3}\delta_{j_{4},3}$ of two Kronecker
deltas guarantees that rings [c] and [d] cannot both be produced by CO
cloud $j=3$, since the term in parentheses vanishes when
$j_{3}=j_{4}=3$.

The likelihood distribution with all three priors in place is shown as
a solid black curve in Fig.~\ref{fig:likelihood}.  The 68\% and 99.7\%
confidence intervals (corresponding to the one- and three-sigma
uncertainties in the distance to Circinus X-1) are plotted as well.

The best fit distance to \cir with 1-sigma statistical
uncertainties\footnote{Uncertainties in eq.~\ref{eq:distance} include
  the effect of random cloud-to-cloud motions {\em and} streaming
  motions as discussed in \ref{sec:distance}. Considering {\em only}
  random cloud-to-cloud motions would reduce the uncertainties to
  $D_{\rm Cir X-1} = 9.4^{+0.6}_{-0.8} \,{\rm kpc}$, in which case an
  allowance for a systematic distance error due to streaming motions
  of order $\pm$0.5 kpc (corresponding to a velocity error of 5km/s)
  should be made.} is
    \begin{equation}
      \label{eq:distance}
      D_{\rm Cir X-1} = 9.4^{+0.8}_{-1.0}
      \,{\rm kpc}
    \end{equation}
    Quantities that depend on $D_{\rm Cir X-1}$ listed in this paper
    (such as the relative cloud distance $x$) are calculated using the
    best fit value from eq.~(\ref{eq:distance}) and are listed in gray
    color in Table \ref{tab:clouds} and Fig.~\ref{fig:clouds}.
    
The distance listed in eq.~(\ref{eq:distance}) is consistent with the
dust-cloud associations of rings [a] and [b] with clouds 2$_{\rm CO}$
and 1$_{\rm CO}$, respectively, it places the cloud responsible for
rings [c] and [d] at velocities associated with CO components
2$_{\rm CO}$ and 3$_{\rm CO}$, respectively, and it places the CO mass
of cloud 4$_{\rm CO}$ in the foreground of all of the scattering
screens, consistent with the observed excess photo-electric absorption
observed in the X-ray images that coincides with the location of cloud
4$_{\rm CO}$.  The locations and velocities to the four scattering
screens derived from the best fit distance are plotted in the top
panel of Fig.~\ref{fig:co_spectra}, showing that the dust clouds fall
close to the Galactic rotation curve.

We overplot the best-fit distance and velocity scale as a function of
observed angle $\theta$ in gray on the top x-axis in
Fig.~\ref{fig:clouds}.  Figure \ref{fig:clouds} also shows the
scattering depth $\left.N_{\rm H}d\sigma/d\Omega\right|_{\rm med}$ in
gray. Because it requires knowledge of $x(D)$, this quantity is
distance dependent.

The distance derived here is both geometric and kinematic, given that
we use kinematic distances to molecular clouds to anchor the {\em
  absolute} distance scale, and more accurate {\em relative} geometric
distances to the scattering screens. As such, one might refer to this
method as either pseudo-kinematic or pseudo-geometric. Given that the
predominant error in our method stems from the kinematic aspect of the
distance determination, we will simply refer to it as a kinematic
distance.

For completeness, we will also list the constraints on
    the distance with only a sub-set of priors 1-3 in place:

\subsubsection{Distance estimate with the two strong priors only}
\label{sec:two_priors}
With both prior 1 and 2 in place, a secondary peak appears at
$7.8\,{\rm kpc}$ (solid gray curve in Fig.~\ref{fig:likelihood}), with
a three-sigma lower limit of $D_{\rm Cir}>6.3\,{\rm kpc}$.  This peak
corresponds to the solutions for which both ring [c] and [d] are
produced by cloud 3$_{\rm CO}$. The formal one-sigma uncertainty of
the distance to \cir from only the two strong priors 1 and 2 is
$D_{\rm Cir X-1} = 9.4^{+0.8}_{-1.7} \,{\rm kpc}$.

\subsubsection{Distance estimate with only prior 1 in place}
\label{sec:one_prior}
With only the first prior in place, the likelihood distribution has
two roughly equally strong peaks, one at $7.4\,{\rm kpc}$ and one at
$9.4\,{\rm kpc}$ (dashed gray curve in Fig.~\ref{fig:likelihood}), and
the possible distance range is restricted from 4.5 to 11 kpc.  The
second strong peak at lower distance corresponds to a solution for
which both rings [a] and [b] are associated with component
1$_{\rm CO}$ and rings [c] and [d] are associated with component
3$_{\rm CO}$.

\subsubsection{Lack of distance constraints without any priors}
\label{sec:no_priors}
Without placing {\em any} priors on dust-CO cloud associations, the
source distance is not formally constrained, as can be seen from the
dotted curve in Fig.~\ref{fig:likelihood}, which has multiple peaks at
distances from 3 to 15 kpc.

\subsection{Implications of a Distance of $9.4\,{\rm kpc}$ for the
  Properties of \cir}
\label{sec:implications}
Our three-sigma lower limit of $D>6.1\,{\rm kpc}$ on the distance to
\cir is inconsistent with previous claims of distances around 4 kpc
\citep{iaria:05}.  On the other hand, the 1-sigma distance range
listed in eq.~(\ref{eq:distance}) is consistent with the distance
range of $7.8-10.5\,{\rm kpc}$ based on the observations of type I
X-ray bursts by \citet{jonker:04}, which further strengthens the case
for the kinematic distance presented here.

A distance of $9.4\,{\rm kpc}$ is only marginally larger than the
fiducial value of $8\,{\rm kpc}$ adopted in \citet{heinz:13}.  The age
of the supernova remnant and thus the age of the X-ray binary
increases slightly to $t_{\rm Cir} \sim 3,000\,{\rm yrs}$, well within
the uncertainties estimated in \citet{heinz:13}, while the supernova
energy would increase by about 40\%.

At a distance of 9.4 kpc, the X-ray binary itself has exceeded the
Eddington luminosity for a 1.4 solar mass neutron star frequently
during the time it has been monitored by both the All Sky Monitor on
the {\rm Rossi X-ray Timing Explorer} and MAXI, most recently during
the 2013 flare that gave rise to the rings reported here.  For
example, during the peak of its bright, persistent outburst between
about 1995 and 1997, the mean 2-10 keV luminosity of the source was
approximately $L_{\rm 1995,2-10keV} \sim 3\times 10^{38}\,{\rm
  ergs\,s^{-1}}\,\left(D/9.4\,{\rm kpc}\right)^{-2}$.  We estimate the
peak flux measured by MAXI during the 2013 flare to be about 2.2
Crabs, corresponding to a peak luminosity of $L_{\rm peak,2-10keV}
\sim 5.6 \times 10^{38}\,{\rm ergs\,s^{-1}}\,\left(D/9.4\,{\rm
    kpc}\right)^{-2}$.

One of the most intriguing properties of \cir is the presence of a
powerful jet seen both in radio \citep{stewart:93} and X-rays
\citep{heinz:07}.  \citet{fender:04c} reported the detection of a
superluminal jet ejection, launched during an X-ray flare. Taken at
face value, the observed proper motion of $\mu \sim 400\,{\rm
  mas\,day^{-1}}$ at a distance of $9.4\,{\rm kpc}$ implies an
apparent signal speed and jet Lorentz factor of $\Gamma \gtrsim
\beta_{\rm app}=v_{\rm app}/c \gtrsim 22\,\left(D/9.4\,{\rm
    kpc}\right)$ and a jet inclination of $\theta_{\rm jet} \lesssim
3^{\circ} \left(9.4\,{\rm kpc}/D\right)$.  This would make the jet in
\cir even more extreme than the parameters presented in
\citet{fender:04c}, who assumed a fiducial distance of $6.5\,{\rm
  kpc}$.

\subsection{The Galactic Positions of Circinus X-1 and Scattering
  Screens [a]-[d]}
\label{sec:position}
At a distance of around 9.4~kpc, \cir is likely located inside the
far-side of the Scutum-Centaurus arm.  A location inside a (star
forming) spiral arm is consistent with the young age of the binary.

The CO cloud nearest to the Sun, component 4$_{\rm CO}$ at $\sim
-33\,{\rm km\,s^{-1}}$, is likely located in the near-side of the
Scutum-Centaurus arm, explaining the large column density responsible
for the foreground absorption.

CO clouds 1$_{\rm CO}$ through 3$_{\rm CO}$ are at intermediate
distances; association of these clouds with either the Norma arm (or
the Scutum-Centaurus arm for cloud 3$_{\rm CO}$) are plausible but
difficult to substantiate.

Section \ref{sec:distance} discussed a possible error due to streaming
motions of CO clouds relative to the LSR as part of the large scale
Galactic structure. For streaming motions to strongly affect the
measured radial velocity to a particular cloud, the cloud would have
to be located near a major spiral arm.

The dominant spiral arm towards \cir is the near side of the
Scutum-Centaurus arm, which crosses the line-of-sight at roughly
$3-4\,{\rm kpc}$, which we tentatively identified with cloud
4$_{\rm CO}$.  CO clouds 1$_{\rm CO}$ and 2$_{\rm CO}$, which dominate
the likelihood distribution plotted Fig.~\ref{fig:likelihood}, are
likely located far away from the center of that spiral arm. Thus,
streaming motions are expected to be moderate (see, e.g.,
\citealt{jones:12} and the artist's impression of Galactic structure
by Benjamin \&
Hurt\footnote{http://solarsystem.nasa.gov/multimedia/gallery/Milky\_Way\_Annotated.jpg}),
which will limit the deviation of the observed radial velocity of
cloud from Galactic rotation. Our error treatment is therefore
conservative in including the full velocity dispersion of streaming
motions as estimated by \citet{mcclure:07} for the fourth quadrant.

Given the Galactic rotation curve plotted in
Fig.~\ref{fig:co_spectra}, the radial velocity corresponding to a
distance of 9.4~kpc at the Galactic position of \cir is
$v_{\rm LSR} = -54\,{\rm km\,s^{-1}}$.  \citet{jonker:07} found a
systemic radial velocity of
$v_{\rm rad,Cir} \sim -26\pm 3\,{\rm km\,s^{-1}}$, which would imply a
1D kick velocity of order
$v_{\rm kick,LOS} \sim 25-35\,{\rm km\,s^{-1}}$, consistent with the
values inferred for high-mass X-ray binaries in the SMC \citep{coe:05}
and theoretical expectations for post-supernova center-of-mass
velocities of high-mass X-ray binaries \citep{brandt:95}.

The potential association of \cir with the CO velocity component
5$_{\rm CO}$ at $v_{\rm LOS} \approx 9\,{\rm km\,s^{-1}}$ suggested by
\citet{hawkes:14} would place \cir at the very far end of the allowed
distance range and require a very large relative velocity of that
cloud with respect to the LSR at more than 30$\,{\rm km\,s^{-1}}$,
both of which would correspond to three-sigma deviations. We therefore
conclude that \cir and cloud 5$_{\rm CO}$ are unassociated.

\subsection{Constraints on the Dust Scattering Cross Section}
\label{sec:cross_section}

The most robust constraints on the scattering cross section, and thus
the properties of the dust responsible for the echo, would be derived
by measuring the angular dependence of the brightness of a single ring
over time. This is because such a measurement removes the large
uncertainty in the total hydrogen column towards a particular cloud,
since it would measure the {\em relative} change in brightness.
Unfortunately, because our observations lack the temporal coverage to
follow the brightness of each ring over a large range in scattering
angle, as done by \citet{tiengo:10}, we cannot probe models of the
dust scattering cross section at the same level of detail.

However, we can still place quantitative constraints on
$d\sigma/d\Omega$, given the scattering depth of the rings shown in
Table \ref{tab:clouds} and estimates of the hydrogen column density in
both H$_{2}$ from CO and in neutral hydrogen from the 21cm data, we
can derive constraints on the dust scattering cross section.  The
scattering angle and the scattering depth depend on $x(\theta,D)$, and
the ring-cloud associations depend on distance as well.

\begin{deluxetable}{llll}
  \tablecaption{Hydrogen column density for rings [a]-[d] from
    $^{12}$CO and 21cm spectra for {\em Chandra} ObsID
    15801 \label{tab:hcolumn}} \tablehead{\colhead{Ring:} &
    \colhead{$N_{H_{2}} {\rm [10^{20} cm^{-2}]}$} & \colhead{$N_{HI}
      {\rm [10^{20} cm^{-2}]}$} & \colhead{$N_{H_{\rm tot}} {\rm
        [10^{20} cm^{-2}]}$}} \startdata
  {[}a] & $5.1^{+5.1}_{-2.6}$ & $17.1^{+17.6}_{-9.4}$ & $27.3^{+20.3}_{-10.7}$\\
  {[}b] & $3.9^{+3.9}_{-2.0}$ & $13.4^{+13.7}_{-7.2}$ & $21.2^{+15.7}_{-8.2}$\\
  {[}c] & $7.7^{+7.7}_{-3.9}$ & $32.4^{+32.9}_{-17.1}$ & $47.9^{+36.3}_{-18.8}$\\
  {[}d] & $7.1^{+7.1}_{-3.6}$ & $20.6^{+20.6}_{-10.3}$ & $34.8^{+25.0}_{-12.5}$\\
  \enddata
\end{deluxetable}

To derive an estimate of the {\em molecular} hydrogen column density,
we used the values of N$_{\rm H_{2}}$ from Table~\ref{tab:hcolumn},
adopting a value of the CO-to-H$_{2}$ conversion factor of $x_{\rm CO}
= 2\times 10^{20}$ \citep{bolatto:13}, allowing for an extra factor of
2 uncertainty in $x_{\rm CO}$.  It should be noted that $x_{\rm CO}$
may vary by up to an order of magnitude within individual clouds
\citep{lee:14}.

We estimated the {\em neutral} hydrogen column density from the 21cm
Southern Galactic Plan Survey, following the same extraction regions
used to derive the integrated CO intensity: We restricted the
extraction region to the {\em Chandra} field of view and weighted the
21cm emission by the {\em Chandra} 3-5 keV surface brightness. We then
fitted Gaussian lines to the velocity components, fixing the velocity
centroids of the components to the CO velocities listed in Table
\ref{tab:co_velocities}.

Because the 21cm components are much broader than the CO components,
clouds 1$_{\rm CO}$ and 2$_{\rm CO}$ are blended in 21cm emission and
we cannot fit them as separate components. To estimate the HI column
density in each component, we divide the total column in component
1$_{\rm HI}$ into two components with a column density ratio given by
the ratio of the flux of the CO components 1$_{\rm CO}$ and 2$_{\rm
  CO}$.  We budgeted an additional factor of 2 uncertainty in the
neutral hydrogen column derived from the 21cm data. The results are
listed in Table~\ref{tab:hcolumn}.

\begin{figure}[t]
  \center\resizebox{\columnwidth}{!}{\includegraphics{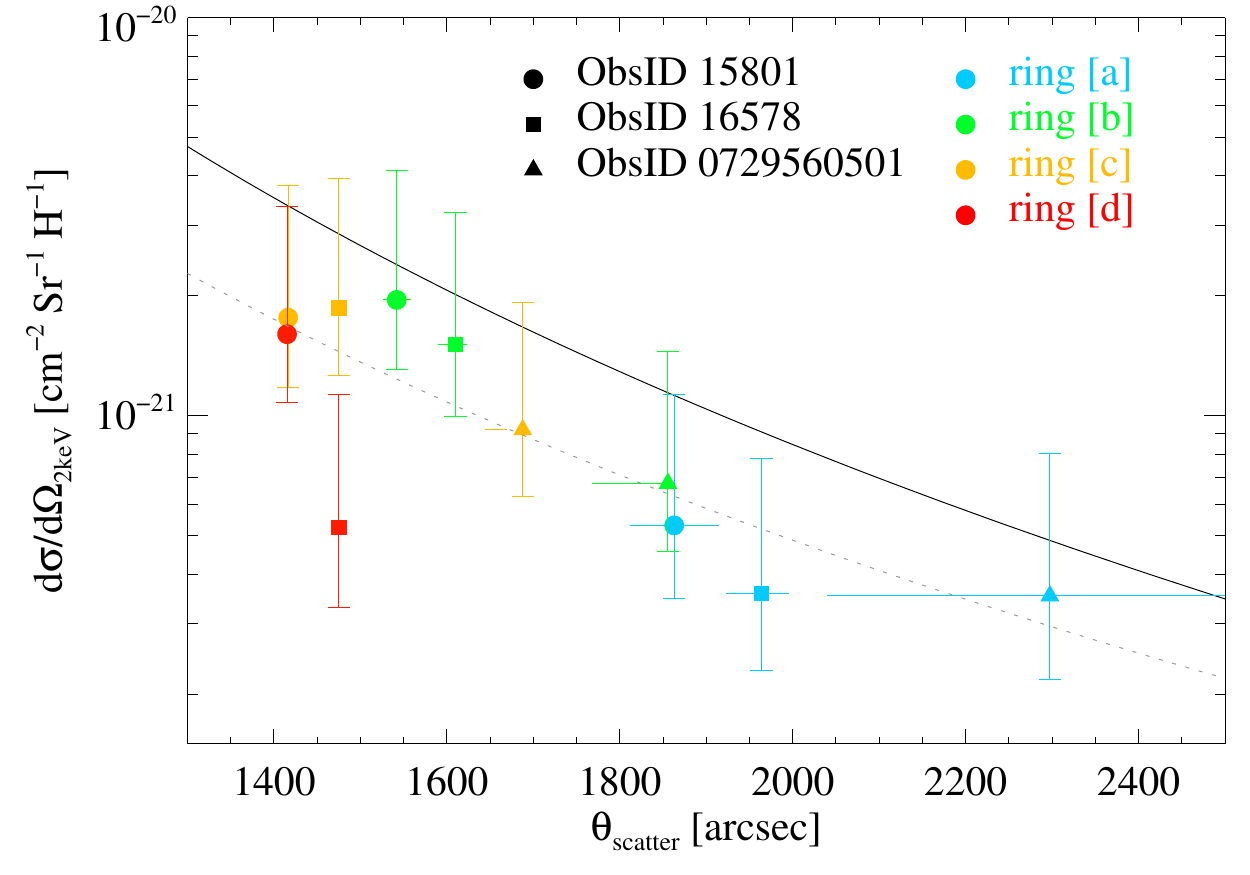}}
  \caption{Dust scattering cross section as a function of scattering
    angle, determined from scattering depth (Table \ref{tab:clouds})
    and the column density from the measured $^{12}$CO intensity and
    21cm HI intensity (Table \ref{tab:hcolumn}), plotted for {\em
      Chandra} ObsID 15801 (circles), ObsID 16578 (squares) and {\em
      XMM} ObsID 0729560501 (triangles) for clouds [a],[b],[c], and
    [d] in green, blue, yellow, and red, respectively, evaluated at
    energy $E=2 keV$. Overplotted is the predicted differential cross
    section from \citet[][solid black line]{draine:03} and the best
    fit powerlaw (dotted line). We do not include ring [d] for {\em
      XMM} ObsID 0729560501 because its centroid falls outside of the
    MOS FOV.\label{fig:cross_section}\vspace*{6pt}}
\end{figure}

Figure \ref{fig:cross_section} shows the differential X-ray scattering
cross section $d\sigma/d\Omega_{\rm sc}$ per hydrogen atom as a
function of scattering angle $\theta_{\rm sc}$, evaluated at 2 keV.
Despite the relatively large scatter, the figure shows the expected
decline with $\theta_{\rm sc}$.  For comparison, the figure shows the
scattering cross section derived by \citet{draine:03}, which agrees
relatively well with the observations in slope and normalization.
Because we associated the entire amount of neutral hydrogen column in
the velocity range $-85$ to $-30\,{\rm km\,s^{-1}}$ to the the four
clouds, the derived cross sections may be regarded as lower limits,
which may be responsible for the vertical offset between model and
data.

We also plot the best-fit powerlaw fit to the cross section as a
function of $\theta_{\rm sc}$ (dotted line):
\begin{equation}
  \left.\frac{d\sigma}{d\Omega}\right|_{\rm 2\,{\rm keV}} \approx
  \left(5.8\pm2.2\right) \times 10^{-21} \left(\frac{\theta_{\rm
        sc}}{1,000'}\right)^{-3.6\pm 0.7}
\end{equation}

It is clear that the large uncertainties in the total hydrogen column
dominate the scatter and uncertainty in this plot. Higher resolution
21cm data woud likely improve the accuracy of this plot
significantly. Ultimately, a direct correlation of the scattering
depth with the absorption optical depth, neither of which depend on
the column density of hydrogen, would provide a much cleaner
measurement of the scattering cross section. Such an analysis is
beyond the scope of this work, however.

\subsection{Predictions for Future Light Echoes and the Prospects for
  X-ray Tomography of the ISM Towards Circinus X-1}
\label{sec:predictions}
The suitable long-term lightcurve and the location in the Galactic
plane with available high-resolution CO data from the Mopra Southern
Galactic Plane Survey make \cir an ideal tool to study the properties
of dust and its association with molecular gas.

Given our preferred distance to Circinus X-1, we can make clear
predictions for observations of future light echoes. Since the four
major rings described in this paper were produced by the molecular
clouds associated with the CO velocity components 1$_{\rm CO}$,
2$_{\rm CO}$, and 3$_{\rm CO}$, future X-ray observations of \cir
light echoes will show similar sets of rings, with radii for a given
time delay determined from eq.~(\ref{eq:deltat}) and Table
\ref{tab:clouds}.

The light echo produced by the CO cloud with the largest column
density (cloud 4$_{\rm CO}$) was already outside of the {\em Chandra}
and {\em XMM} FOVs at the time of the observations. In order to
confirm the dust-to-CO-cloud association presented in this paper,
future X-ray observations of light echoes should include {\em Chandra}
or {\em Swift} observations sufficiently close in time to the X-ray
flare to observe the very bright light echo expected from cloud
4$_{\rm CO}$, which we will refer to as the putative ring [e].  Ring
[e] will be brightest in the Southern half of the image.  Because \cir
exhibits sporadic flares several times per year, it should be possible
to observe this echo with modest exposure times and/or for moderate
X-ray fluence of the flare, given the expected large intensity of the
echo due to the large column density of cloud 4$_{\rm CO}$. 

A detection of ring [e] would improve the fidelity of the distance
estimate primarily because it would test the ring-cloud-associations
assumed in our three priors. However, the statistical improvement on
the distance estimate would be marginal because the low relative
distance $x_{4_{\rm CO}} \sim 0.25$ to cloud 4$_{\rm CO}$ implies a
large change in $x_{4_{\rm CO}}$ with $D$ in such a way that
$D_{4_{\rm CO}}=x_{4_{\rm CO}}D$ (and thus also the velocity
constraint on cloud 4$_{\rm CO}$) is rather insensitive to $D$.

Future observations at smaller delay times will also provide
constraints on the scattering cross section at smaller scattering
angles.  Constraints on $d\sigma/d\Omega$ from {\em single} rings at
{\em different} scattering angles are inherently less error-prone, as
the error in the normalization of $N_{\rm H}$ affects all measurement
equally, probing the {\em relative} change of scattering cross section
with $\theta_{\rm sc}$.

Better temporal coverage of the light echo from earlier times will
allow detailed maps of each cloud both from the spatial distribution
of the scattering signal and from the foreground absorption caused by
each of the dust clouds, which can be compared to the scattering depth
and the hydrogen column density from CO and HI measurements.

Should a high-accuracy distance determination to \cir become possible
from other methods, (such as maser parallax to one of the CO clouds,
in particular cloud 4$_{\rm CO}$, which has the highest column
density, or VLBI parallax to the source itself), the distance to {\em
  all} dust components will be determined to within 5\% accuracy or
better (set by the relative angular width $\sigma_{\rm
  theta}/\theta_{\rm med}$ of the dust cloud from Table
\ref{tab:clouds}).  This would allow for extremely accurate tomography
of the dust and gas distribution towards \cir from the existing
observations presented here and any future light echo observations,
providing a powerful tool to study Galactic structure in the fourth
quadrant.

\section{Conclusions}
\label{sec:conclusions}
The giant flare and the associated X-ray light echo presented in this
paper are the brightest and largest set of light-echo rings observed
from an X-ray binary to date. The deviation from spherical symmetry
allows us to make an unambiguous determination of the clouds
responsible for part of the light echo from an X-ray binary.

Because of the full temporal coverage of the lightcurve of the X-ray
flare by MAXI, we are able to reconstruct the dust distribution
towards \cir by deconvolving the radial profile with the dust
scattering kernel derived from the X-ray lightcurve.

The kinematic distance of about 9.4 kpc we derived from the
association of ring [a] with cloud 2$_{\rm CO}$ and ring [b] with
cloud 1$_{\rm CO}$ is consistent with the estimate by
\citet{jonker:04} based on the luminosity of the type I X-ray bursts
of the source \citep{tennant:86}.  It places a large cloud of
molecular gas (cloud 4$_{\rm CO}$ and the associated large absorption
column in the Southern quadrant of the FOV) in the foreground of the
dust screens detected in the light echo.

At a distance of about $9.4\,{\rm kpc}$, the source has exceeded the
Eddington limit for a 1.4 $M_{\odot}$ neutron star frequently during
the time it has been monitored by the All Sky Monitors on the {\rm
  Rossi X-ray Timing Explorer} and MAXI, both during recent outbursts
and, more importantly, in average luminosity during its persistent
high-state in the 1990s.

Taken at face value, the observed pattern speeds in the jet
\citep{fender:04} indicate a very fast jet speed of $\Gamma \gtrsim
22$ and a correspondingly small viewing angle of $\theta_{\rm jet}
\lesssim 2.6^{\circ}$.

Future measurements of light echoes from \cir will allow accurate
measurements of the dust scattering cross section as a function of
scattering angle and an estimation of the dust-to-gas ratio in the
different clouds identified in the field of view towards Circinus X-1.

The technique used to derive the kinematic distance to \cir holds
significant promise for future observations of light echoes from X-ray
transients.

\acknowledgements{We would like to thanks Snezana Stanimirovic, Bob
  Benjamin, Audra Hernandez, Min-Yong Lee, and Eugene Churazov for
  helpful discussions. We would also like to thank the CXC team for
  outstanding support in scheduling and analyzing the {\em Chandra}
  observations.  This research has made use of data obtained from the
  Chandra Data Archive and the {\em Chandra} Source Catalog, and
  software provided by the Chandra X-ray Center (CXC) in the
  application packages {\tt CIAO}, {\tt ChIPS}, and {\tt Sherpa}.
  Support for this work was provided by the National Aeronautics and
  Space Administration through Chandra Award Number GO4-15049X issued
  by the {\em Chandra} X-ray Observatory Center, which is operated by
  the Smithsonian Astrophysical Observatory for and on behalf of the
  National Aeronautics Space Administration under contract
  NAS8-03060. XMM data used in this manuscript were obtained through
  generously scheduled Director's Discretionary Time. This work is
  based on observations obtained with XMM-Newton, an ESA science
  mission with instruments and contributions directly funded by ESA
  Member States and the USA (NASA). During the preparation of this
  manuscripts, we made use of the XMM-{\em ESAS} package. We would
  like to thank the Swift scheduling team for help in planning and
  executing this program.  The CO data was obtained using the Mopra
  radio telescope, a part of the Australia Telescope National Facility
  which is funded by the Commonwealth of Australia for operation as a
  National Facility managed by CSIRO. The University of New South
  Wales (UNSW) digital filter bank (the UNSW-MOPS) used for the
  observations with Mopra was provided with support from the
  Australian Research Council (ARC), UNSW, Sydney and Monash
  Universities, as well as the CSIRO. This research has made use of
  the MAXI data provided by RIKEN, JAXA and the MAXI team. We
  acknowledge the use of public data from the Swift data archive.}

\end{document}